\documentclass[prd,superscriptaddress,amsfonts,amssymb,amsmath,showpacs,twocolumn,showkeys]{revtex4-1}
\usepackage{amsmath}
\usepackage{epsfig}
\usepackage{graphics}
\usepackage[colorinlistoftodos]{todonotes}
\usepackage{color}
\usepackage{xcolor}
\usepackage{graphicx}
\usepackage[font={footnotesize,it}]{caption}
\usepackage[colorlinks=true,
            linkcolor=black,
            urlcolor=blue,
            citecolor=blue]{hyperref}
    
\def\be{\begin{equation}} 
\def\ee{\end{equation}}
\def\bea{\begin{eqnarray}}
\def\eea{\end{eqnarray}}

\begin{document}

\title{Impact of the Rastall parameter on perfect fluid spheres}

\author{Sudan Hansraj}
\email{hansrajs@ukzn.ac.za}
\affiliation{Astrophysics and Cosmology Research Unit, University of KwaZulu Natal, Private Bag X54001, Durban 4000,
South Africa.}

\author{Ayan Banerjee}
\email{ayan_7575@yahoo.co.in}
\affiliation{Astrophysics and Cosmology Research Unit, University of KwaZulu Natal, Private Bag X54001, Durban 4000,
South Africa.}

\author{Phongpichit Channuie}
\email{channuie@gmail.com}
\affiliation{School of Science, Walailak University,
Nakhon Si Thammarat, 80160 Thailand.}

\begin{abstract}

We examine the effects of the Rastall parameter on the behaviour of spherically symmetric static distributions of perfect fluid matter. It was claimed by Visser [Physics Letters B, 782, 83, (2018)] that the Rastall proposition is completely equivalent to the Einstein theory. While many authors have raised contrary arguments, our intention is to analyze the properties of Rastall gravity through variation of the Rastall parameter in the context of perfect fluids spheres that may be used to model neutron stars or cold fluid planets. This analysis also serves to counter the claim that Rastall gravity is equivalent to the standard Einstein theory.  It turns out that the condition of pressure isotropy is exactly the same as for Einstein gravity and hence that any known solution of the Einstein equations may be used to study the effects of the Rastall dynamical quantities. Moreover, by choosing the well studied Tolman metrics, we discover that in the majority of cases there is substantial deviation from the Einstein case when the Rastall parameter vanishes and in cases where the Einstein model displays defective behaviour, certain Rastall models obey the well known elementary requirements for physical plausibility. These empirical findings do not support the idea that Rastall theory is equivalent to Einstein theory as several deviations in physical behavior are displayed as counter-examples.

\end{abstract}

\keywords{xxxxxxxxxxxxxxx}
\pacs{xxxxxxxxxxxxxx}

\maketitle

\section{Introduction}

The accelerated expansion of the universe plays an important role in the dynamical history of our universe. There exists strong evidence that the universe has passed through an inflationary phase at early times and there has been increasing substantiation of the late-time cosmic acceleration through a large variety of observational data. Some of these include the Type-Ia supernovae \cite{Perlmutter}, Baryon Acoustic Oscillations \cite{Eisenstein} and the 
Wilkinson Microwave Anisotropy Probe (WMAP) in the Cosmic Microwave Background (CMB) \cite{Spergel}. What mysterious force is actually responsible for the acceleration remains an open and tantalizing question. To date, there is no competing model of the standard theory that has adequately addressed this problem and this situation motivates the requirement for alternatives. Many suggestions have been made, among which the $\Lambda$-cold-dark-matter ($\Lambda$CDM) model has been widely applied to the interpretation of a range of cosmologically-oriented observations. 
Following this philosophy, several other models have been proposed to incorporate the  cosmic acceleration, namely, generalizations of the Chaplygin gas, quintessence fields and so-called tachyon models. A simple way to parameterize the dark energy model is to consider an equation of state (EOS)  $\omega =p/\rho$; where $p$ is the spatially homogeneous pressure and $\rho$ is the dark energy density.
 
Another possibility requires the generalization of Einstein's theory of gravity, that is, 
one starts from the curvature description of gravity. Mainly, modifications 
of gravity include models where the Einstein-Hilbert Lagrangian
is supplemented with additional curvature terms $(f(R, R_{\mu\nu}R^{\mu\nu}, R_{\mu\nu\alpha\beta}R^{\mu\nu\alpha\beta}, G,....)$. In particular,
 the Gauss-Bonnet \cite{Nojiri,Nojiri1} invariant $G=R^2-4R_{\mu\nu}R^{\mu\nu}+R_{\mu\nu\alpha\beta}R^{\mu\nu\alpha\beta}$ and is a special case of the more general Lovelock polynomials \cite{NKD,dad-hans}. In this context, a class of theories, called $f(R)$-gravity, where $f(R)$ is a generic 
function of the Ricci scalar $R$, plays an important role as a modification of the Einstein-Hilbert  gravitational Lagrangian density, first considered by \cite{Buchdahl1}. As a result, it attracted serious attention because of its greater geometrical degrees of freedom instead of searching for new material ingredients, and is further developed in \cite{Cardone,Capozziello2,Capozziello3}. Furthermore, there are also different classes of modified gravity theories such as so-called $f(R,T)$ theories of gravity, where $R$ is the Ricci scalar and $T$ is the trace of the energy-momentum tensor have provided a number of extremely interesting results on  both cosmological \cite{Moraes,Moraes1,Baffou,Shabani,Myrzakulov,Jamil,Sharif,Alvarenga,ovalle}
 and astrophysical \cite{Carvalho,Das1,Bhatti} scales. These proposals are considered as phenomenological motivations in terms of compelling consequences, such as a violation of one or more energy conditions, incompatible with the Newtonian regime.

From a  mathematical point of view, modifications of gravity were related to
represent  the gravitational field behaviour near curvature singularities and possible to create some first order approximation for the quantum theory of gravitational fields. In the curvature-matter gravity theory, when the covariant divergence of the energy-momentum  tensor is non-zero, matter and geometry fields are coupled to each other in a non-minimal way. In this connection a specific application to the modification of Einstein's theory was proposed by P. Rastall in 1972 \cite{Rastall,Rastall1}, where the  covariant divergence of energy-momentum tensor proportional to the covariant divergence of the curvature scalar, i.e., $T^{\mu}_{\nu;\mu}$ $\propto$ $R_{;\nu}$.
As a result, the non-minimal coupling leads to a modified general relativity theory. It should be mentioned here that Rastall's model has been initially motivated to  challenge the energy-momentum conservation law in the curved spacetime without  violating the Bianchi identities.  Hence, in the weak field approximation,  the usual expressions are preserved.

Also, an important ingredient of the Rastall theory is that the field equations 
are simpler than those of other curvature-matter theories and are consequently easier to investigate. Additionally, in confronting the data, Batista \emph{et al.} \cite{Batista1} studied cosmological scenarios based on the Rastall proposal (Rastall Cosmology) with the assumption of a two-fluid model, one component represents vacuum energy whereas the other pressure-less matter (dust).
In order to establish such a new gravitational theory, it is important to study the construction of  astrophysical  models also. For example,  by special setting of the parameters of the  Rastall theory, particular rotating and non-black hole solutions were conducted \cite{Heydarzade,Kumar,Heydarzade1}. The structure of neutron stars in Rastall modification of gravity was investigated in refs \cite{Oliveira}.
Recently, Moradpour and Salako \cite{Moradpour} studied thermodynamic properties 
of static spherically symmetric field equations in Rastall theory and also in 
flat FLRW universe \cite{Moradpour1}.  A study regarding
the asymptotically flat traversable wormhole solutions has been studied in Ref. \cite{Moradpour2}.

The exact analytic solutions of the Einstein field equations in GR have provided a number of important results ranging from singularity free interior solutions to the physical understanding of relativistic phenomena. The physical processes are governed by sets of physical variables yielding an outnumber of the independent field equations. For this reason many investigators have pursued a 
 variety of techniques to obtain exact solutions. To date the number of solutions is large 
\cite{Kramer} and growing, but among them only few solutions are physically valid and satisfy some basic requirements such as regularity within the interior, existence of a vanishing pressure  boundary surface, energy conditions, causality, etc. The review written by Delgaty and Lake \cite{Delgaty} of
over 130 solutions reveals that only nine could be classified as physically relevances satisfying 
elementary physical conditions.

An important motivation for this empirical investigation is to question the claim of Visser \cite{Visser} that the 45 year old  Rastall theory 
of gravity is nothing more than the Einstein equations in different garb. In Rastall theory the divergence of the energy momentum tensor is proportional to the derivative of the Ricci scalar. In GR this divergence vanishes by the Ostrogradsky's theorem. It is therefore difficult to see mathematically how the derivative of the general Ricci scalar can be diffeomorphic to zero in general. Visser demonstrates that the geometrical part of the field equations are identical. There is no issue with this conclusion and we show in this article that the equation of pressure isotropy for Rastall theory and general relativity are identical meaning that the metric potentials for both theories are identical. Then Visser asserts that the energy momentum tensor representing the matter sector is composed of a modified conserved quantity which is essentially split up in two parts each of which is not independently conserved. It is clear that the modified energy momentum tensor does not correspond to a classical perfect fluid. In addition Visser does not offer any experimental method to determine which energy momentum tensor is the physically correct one. 
The matter has been adequately dealt with by  Darabi {\it {et al}} \cite{Darabi} who have shown that the incorrect energy momentum tensor was alluded to by Visser. In the present article we consider, as a simple counter-example,  the famous Schwarzschild interior metric and show that while this metric yields an incompressible fluid sphere (constant density) in general relativity, the model in Rastall theory using the exact same metric does not display constant density but is a variable quantity dependent on the radial value. The question of what metric generates a constant density fluid in the Rastall theory remains open. The governing system of nonlinear differential equations have proved to be intractable at this point in time.  In this paper, we are going to study the behaviour of well known stellar models 
obtained by Tolman \cite{Tolman} within the context of Rastall theory with a view to showing how these models deviate from general relativity. Indeed in not a few cases, we find that the Rastall model displays features consistent with physical reality as opposed to its counterpart general relativity.

The plan of the paper is as follows: We introduce a brief review of Rastall gravity in Section II. We derive the 
field equation evolution of perfect fluid matter in Rastall theory in Section III.
In the next section, after referring to the Rastall theory we study all Tolman solutions, and examine the dynamical properties between Einstein theory and Rastall theory in Section IV. Finally we conclude in the last section.

\section{ The Rastall theory of gravity}

Here, we start from the short introduction to Rastall theory of gravity, 
which was introduced by P. Rastall \cite{Rastall,Rastall1}.  The starting point
of this hypothesis lying on the fact that $T^{ab}_{;b}\neq 0$, i.e.,
the usual conservation law of the energy momentum tensor does not hold. 
Based on the Rastall's theory the energy momentum tensor can be determined as
\begin{equation}
T^{\mu\nu}_{;\mu} = \alpha R^{;\nu}, \label{1}
\end{equation}
where $R$ is Ricci scalar, and the Rastall parameter $\alpha$ 
which quantifies the deviation from the  Einstein theory of General Relativity (GR).
Thus, a non-minimal coupling of matter fields to geometry is considered such that the 
usual conservation law is recovered in the flat spacetime, which leads to the 
modification of Einstein's tensor as 
\begin{eqnarray}
G_{\mu\nu}+ \gamma g_{\mu\nu}R= \kappa Tg_{ab}, \label{2}
\end{eqnarray}
where $\gamma = k \alpha$ and $k$ is the Rastall gravitational coupling
constant. Eventually, one can express the above equation in the following form 
\begin{eqnarray}
G_{\mu\nu}= \kappa T_{\mu\nu}^{\text{eff}}, \label{3}
\end{eqnarray}
where $T_{\mu\nu}^{\text{eff}}$ is the effective energy-momentum tensor defined as
\begin{eqnarray} \label{4}
T_{\mu\nu}^{\text{eff}}= T_{\mu\nu}-\frac{\gamma T}{4\gamma -1}g_{\mu\nu}.
\end{eqnarray}
The expression for $T_{\mu\nu}^{\text{eff}}$ is given by \cite{Moradpour2}
\begin{eqnarray}
S^0_0 \equiv -\rho^{\text{eff}}  =-\frac{(3\gamma-1)\rho+\gamma(p_r+2p_t)}{4\gamma-1},\\\label{5}
S^1_1 \equiv p^{\text{eff}}_r  = \frac{(3\gamma-1)p_r+\gamma(\rho-2p_t)}{4\gamma-1},\\\label{6}
S^2_2= S^3_3\equiv p^{\text{eff}}_t  = \frac{(2\gamma-1)p_t+\gamma(\rho-p_r)}{4\gamma-1},\label{7}
\end{eqnarray}
where $\rho$ is the energy density, $p_r$ and  $p_t$ are the radial and tangential pressures, 
respectively which are in general different ($p_r \neq  p_t$). It is to be noted that the energy-momentum 
tensor is conserved when $\alpha \rightarrow 0$ as in the case of general relativity.
Also, for a traceless energy-momentum source, such as the electromagnetic source, the Eq.~(\ref{3}), leads 
to $T_{\mu\nu}^{\text{eff}}=T_{\mu\nu}$, and it benefits from the fact that standard 
Einstein gravity is again recovered. In this regard, the Einstein 
solutions for $T = 0$, or equivalently $R = 0$, are also valid in the Rastall theory of $\kappa$.
Another aspect which we should note from the definitions of the Newtonian limit is 
that  the Rastall parameter $\alpha$ and gravitational coupling constant $\kappa$   diverge at  $\gamma =1/4$ and $\gamma =1/6$, respectively, which does not conform to physical reality \cite{Moradpour2}.

Indeed, it is more commonly assumed that  $\kappa =1$, then $\gamma=\alpha=1$, which 
is what we are assuming here in this system of units. Now that the units have been
clarified  the Rastall field Eqs.~(\ref{3}) and (\ref{4}), may be written as 
\begin{eqnarray}
G_{\mu \nu}= T_{\mu\nu}-\frac{\alpha T}{4\alpha -1}g_{\mu\nu}, \label{8}
\end{eqnarray}
Note that when one sets  $\alpha$ to zero, one gets the original TOV equations for general relativistic quantities.

\section{FIELD EQUATIONS}

In order to facilitate a direct comparison with the work
of Tolman, we follow his conventions. We begin with the static 
spherically symmetric metric in Schwarzschild-like coordinates (t, r, $\theta$, $\phi$), 
which is given by the following line element
\begin{equation}
ds^{2} = -e^{\nu(r) } \, dt^{2}+e^{\lambda(r)} dr^{2} +r^{2}(d\theta ^{2} +\sin ^{2} \theta \, d\phi ^{2}), \label{9}
\end{equation}
where the gravitational potentials $\nu$ and $\lambda$ are functions
of the radial coordinate $r$ only. We utilise a comoving fluid
4-velocity $ u^a = e^{-\nu /2} \delta_0^a $ and consider a perfect fluid source
with energy momentum tensor $T_{\mu\nu} = (\rho + p)u_{\mu} u_{\nu} + p g_{\mu\nu}$, 
where the Greek indices $\mu$ and $\nu$ run from 0 to 3. Here the 
quantity $\rho$ is the energy-density and $p$ is the isotropic pressure, respectively.

Substituting the
non-vanishing trace part of the total energy-momentum tensor into Eq.~(\ref{4}), 
and re-organizing the resulting terms we end up with the following effective set of   the Rastall field equations 
\begin{widetext}
\begin{eqnarray}
\frac{(4\alpha - 1)e^{-\lambda}}{r^2} \left(1 - r\lambda' + e^{\lambda}\right) = - 3\alpha p - (3\alpha -1)\rho ,  \label{6a} \\
\frac{(4\alpha - 1)e^{-\lambda}}{r^2} \left(1 + r\nu' - e^{\lambda}\right) = (\alpha - 1) p + \alpha \rho ,  \label{6b}\\
r^2(2\nu'' + \nu'^2   + \nu'\lambda' ) - 2r(\nu' +\lambda')  + 4(e^{\lambda} - 1) = 0. \label{6c}
\end{eqnarray}
\end{widetext}
where Eq.(\ref{6c}) is the equation of pressure isotropy identical to that of standard Einstein theory. This means that any of the roughly 120 exact solutions reported in the literature may be used to study the Rastall theory of gravity in the case of compact objects. The energy density and pessure may be expressed independently as
\begin{eqnarray}
\rho &=& \frac{e^{-\lambda}}{r^2 } \left(-\lambda_{1} - (\alpha -1)r \lambda' + 3\alpha r \nu'  \right), \label{7a}  \\
p &=& \frac{e^{-\lambda}}{r^2 } \left(\lambda_{1}
+ \alpha r \lambda' - (3\alpha - 1)r \nu'   \right). \label{7b} 
\end{eqnarray}
where we have defined a new variable $\lambda_{1}=(4\alpha - 1)(e^{\lambda} - 1)$. Note that the inertial mass density $\rho + p$  is given by 
\begin{eqnarray} \rho  +  p = \frac{e^{-\lambda} (\nu' + \lambda')}{r}
\end{eqnarray}
which is independent of the Rastall parameter, $\alpha$.

\section{The Tolman Solutions}

In the present study we have three independent field equations mentioned above, with 
four unknowns $\lambda$, $\nu$, $\rho$ and $p$, as functions of $r$ as in the standard theory. A promising avenue to solve the system of equations, Tolman \cite{Tolman} developed a method to obtain an exact analytic solution to the spherically symmetric, static Einstein equations with a perfect fluid source, in terms of known analytic functions. Specifically, we proceed the same approach, and 
derive all the Tolman Solutions in Rastall gravity, in order to compare with general relativity. The entire analysis has been performed to examine the behaviour of the energy density, pressure, velocity of sound  and figure out possible the mass profile. Note also that it is customary to neglect the cosmological  constant in astrophysical scales.

\subsection{Tolman I metric (Einstein Universe)}

Tolman commenced with Einstein's assumption of a constant temporal
potential: $e^{\nu}=\text{const.} = c^2$ for some constant $c$.  The remaining metric potential is then found to be
$e^{\lambda}=\frac{1}{1-\frac{r^2}{R^2}}$ where $R$ is another constant. Accordingly the dynamical quantities show $\rho=\frac{3}{R^2}$ and $ p=-\frac{1}{R^2}$.

In Rastall gravity  the density and pressure are calculated as $\rho = \frac{3-6 \alpha }{R^2} $ and $
 p =  \frac{6 \alpha -1}{R^2} $. Positivity of both demands
 $\frac{1}{6} < \alpha < \frac{1}{2}$.
 The energy  conditions assume the forms
 $ \rho - p  =  \frac{4-12 \alpha }{R^2}$,  $\rho + p  =
 \frac{2}{R^2}$ and $  \rho + 3p  =  \frac{12 \alpha }{R^2} $.
 Ensuring these quantities remain positive yields $\alpha$ as $\frac{1}{6} <
 \alpha < \frac{1}{3}$. The mass function is then determined as $ M = \frac{(1-2 \alpha )
 r^3}{R^2}$  and its positivity is guaranteed through $\alpha <
 \frac{1}{2}$. In view of the unrealistic constant density and
 pressure, there is little value in studying this case further.

\subsection{Tolman II metric (Schwarzschild--de Sitter)}

With the prescription $e^{-\lambda-\nu}=\text{constant}$ Tolman
obtained the potentials $
e^{\lambda}=\left(1-\frac{2m}{r}-\frac{r^2}{R^2}\right)^{-1} $ and $
e^{\nu}= c^2 \left(1-\frac{2m}{r}-\frac{r^2}{R^2}\right).$ The
density and pressure emerge as $ \rho=\frac{3}{R^2}$ and $
p=-\frac{3}{R^2}.$ The equation of state $\rho + p = 0$ is evident
and this is characteristic of dark energy models.

The situation in Rastall theory is similar. The density and pressure
are found to be $ \rho = \frac{3-12 \alpha }{R^2}$ and $ p =
\frac{3( 4\alpha -1)}{R^2}$ respectively. The weak and dominant
energy conditions yield $
 \rho - p  =  \frac{6(1-4 \alpha) }{R^2}$ and $\rho + 3p = \frac{6( 4\alpha-1) }{R^2}
 $. Clearly both cannot be simultaneously positive so there is a
 violation of the basic energy conditions. The equation of state
 $\rho + p = 0$ is still valid. The sound speed is meaningless and
 the mass profile obeys $ M = \frac{(1-4 \alpha ) r^3}{R^2}$. A positive mass requires
 $\alpha < \frac{1}{4}$. However, this causes a negative pressure
 hence this case is not feasible in Rastall theory. In the standard
 Einstein theory it represents the gravitational field exterior to a
 spherical body with a cosmological constant.

\subsection{Tolman III metric: Schwarzschild Interior}

Invoking the ansatz $e^{-\lambda}=1-\frac{r^2}{R^2}$ Tolman obtained the temporal potential as $e^\nu =
\left[A-B\left(1-\frac{r^2}{R^2}\right)^{\frac{1}{2}}\right]^2 $. In the Rastall framework, the density and pressure have the forms
\begin{eqnarray}
 \rho &=& \zeta^{-1}\Big(3 (1-2 \alpha ) A R^2 \sqrt{1-\frac{r^2}{R^2}}\nonumber\\&&\quad\quad-3 (4 \alpha -1) B \left(r^2-R^2\right)\Big), \label{30a} \\
 p &=& \zeta^{-1}\Big((6 \alpha -1) A R^2 \sqrt{1-\frac{r^2}{R^2}}\nonumber\\&&\quad\quad+3 (4 \alpha -1) B \left(r^2-R^2\right)\Big), \label{30b}
\end{eqnarray}
where we have defined $\zeta=A R^4 \sqrt{1-\frac{r^2}{R^2}}+B R^2 \left(r^2-R^2\right)$. Observe that the energy density  does not yield the compressible (constant density) fluid sphere as is the case in standard Einstein theory. This demonstrates a deviation of Rastall theory from the standard general relativity. The expressions governing the energy  conditions are given by
\begin{eqnarray}
 \rho - p & = & \zeta_{1}^{-1}\Big(4 (1-3 \alpha ) A R^2 \sqrt{1-\frac{r^2}{R^2}}\nonumber\\&&\quad\quad-6 (4 \alpha -1) B \left(r^2-R^2\right)\Big), \label{31a}\\
 \rho + p & = & \zeta_{2}^{-1}\Big(2 A \sqrt{1-\frac{r^2}{R^2}}\Big),  \label{31b}\\
\rho + 3p &=& \zeta_{1}^{-1}\Big(12 \alpha  A R^2 \sqrt{1-\frac{r^2}{R^2}}\nonumber\\&&\quad\quad+6 (4 \alpha -1) B \left(r^2-R^2\right)\Big). \label{31c}
\end{eqnarray}
where we have defined new variables: $\zeta_{1}=A R^4 \sqrt{1-\frac{r^2}{R^2}}+B R^2 \left(r^2-R^2\right)$ and $\zeta_{2}=A R^2 \sqrt{1-\frac{r^2}{R^2}}+B \left(r^2-R^2\right)$. The square of  sound-speed evaluates to a constant
 \be
 \frac{dp}{d\rho} =\frac{1}{3 \alpha }-1, \label{32}
 \ee
 and constrains $\alpha$ to $\frac{1}{6} < \alpha < \frac{1}{3}$. Note that in general relativity the sound speed is infinite in view of the constant density, however, in contrast the Rastall theory supports a subluminal fluid congruence. The gravitational mass has the form
\begin{widetext}
\begin{eqnarray}
 m &=& 3 \left(\frac{2 \alpha  A^2 r}{B^2}+\frac{2 \alpha  A^2 R^2 \beta_3}{B^3 \sqrt{r^2-R^2}}
  + \frac{\alpha  A R \Pi_3}{B^3}
 -\frac{2 \alpha  A^2 R \sqrt{A^2-B^2} \gamma_3}{B^3} 
 + \frac{\alpha  A r \sqrt{1-\frac{r^2}{R^2}}}{B} - \frac{(4 \alpha -1) r^3}{3 R^2}\right). \label{33}
\end{eqnarray}
Note we have used the notations
\end{widetext} 
 $ \beta_3=\sqrt{A^2-B^2} \sqrt{1-\frac{r^2}{R^2}} \tanh ^{-1}\left(\frac{A r}{\sqrt{A^2-B^2} \sqrt{r^2-R^2}}\right)$, 
 $\gamma_3 = \tan ^{-1}\left(\frac{B r}{R \sqrt{A^2-B^2}}\right)$, and $\Pi_3=\left(2 A^2-B^2\right) \sin ^{-1}\left(\frac{r}{R}\right)$ for simplicity. We observe the richer behavior in the Rastall version of the physical
quantities compared to the Einstein one. In particular, note
that the density is not constant in general for these metric
potentials compared with the standard theory. The results of the standard case
are regained for  $\alpha = 0$. For a comparative analysis of the
impact of the Rastall parameter, we make plots for graphical illustrations.
Throughout this work, a thick solid line represents the Einstein
case, while the other curves correspond to different values of
$\alpha$ as follows: dotted $(\alpha = 0.25)$, dashed $(\alpha =
0.5)$, dotted-dashed $(\alpha = 2)$ and thin line ($\alpha = - 2$).
The question of what metric generates a constant density fluid
sphere in Rastall gravity is still open and will be addressed in a
different article.

\begin{figure}
    \centering
   \includegraphics[width=0.3\textwidth, height=0.24\textheight]{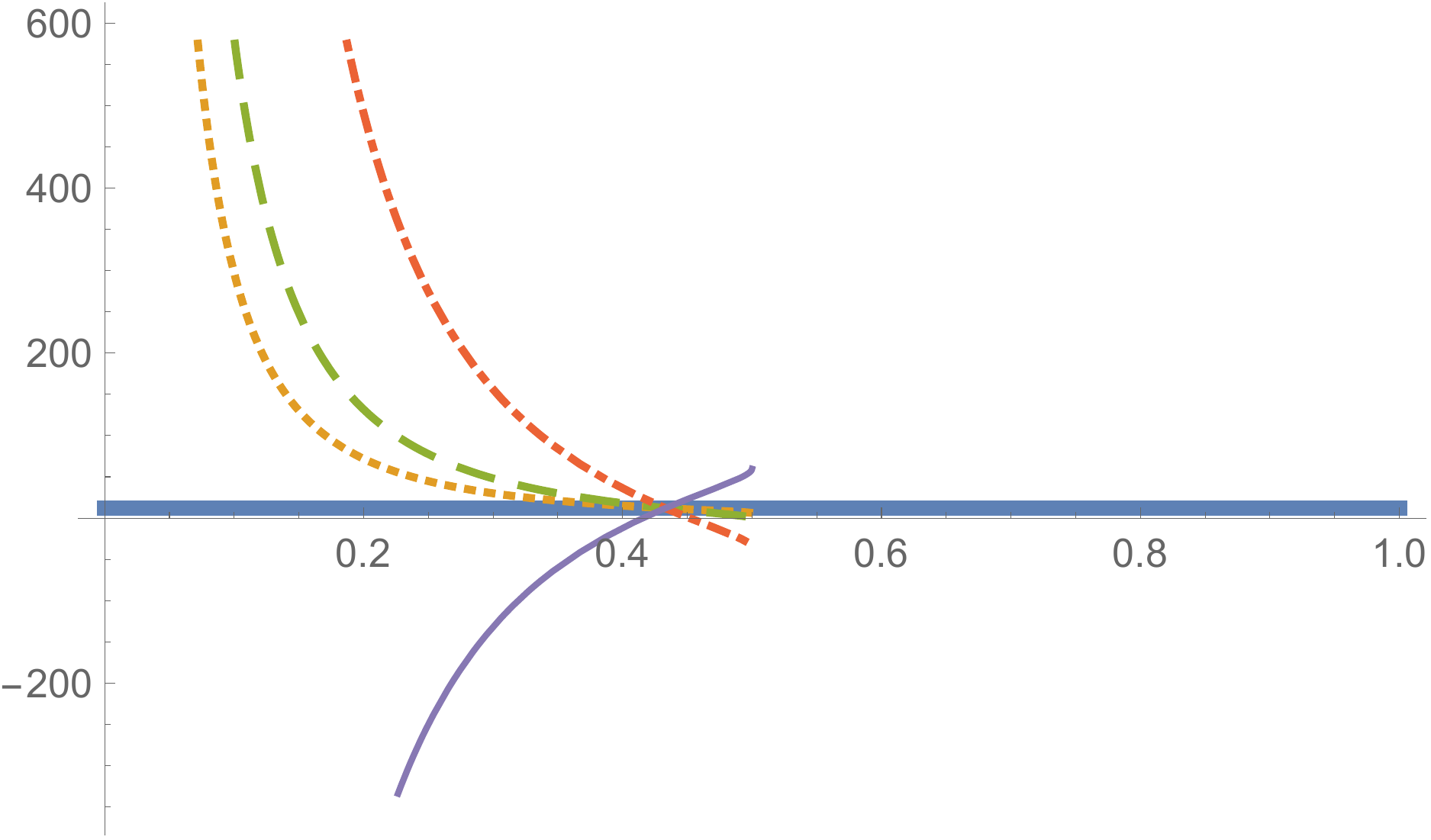}
    \caption{Plot of energy density ($\rho$) versus radius ($r$): Tolman III}
\end{figure}

\newpage
\begin{figure}
    \centering
  \includegraphics[width=0.3\textwidth, height=0.24\textheight]{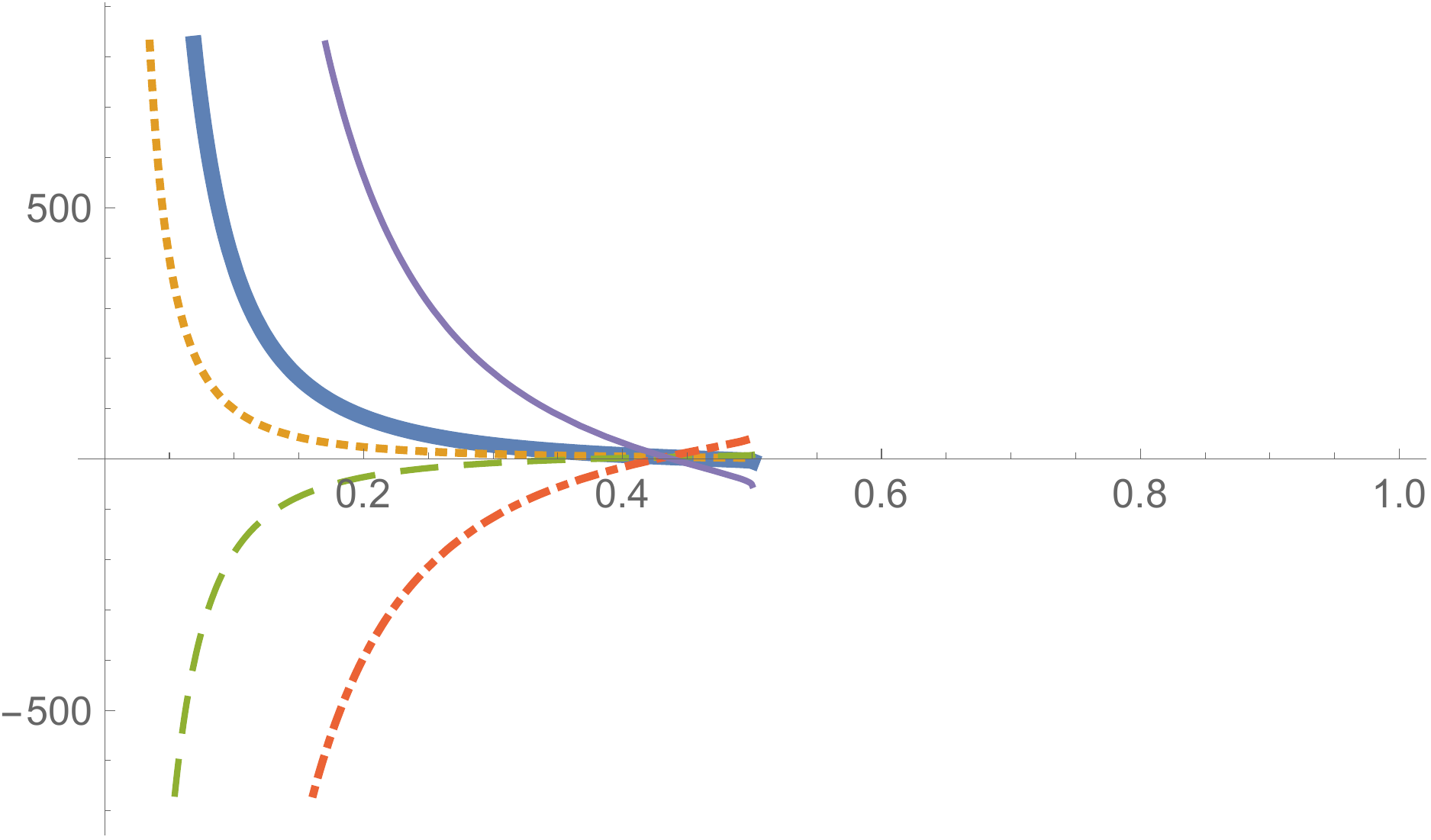}
    \caption{Plot of pressure ($p$) versus radius ($r$): Tolman III}
\end{figure}

\newpage
\begin{figure}
    \centering
    \includegraphics[width=0.3\textwidth, height=0.23\textheight]{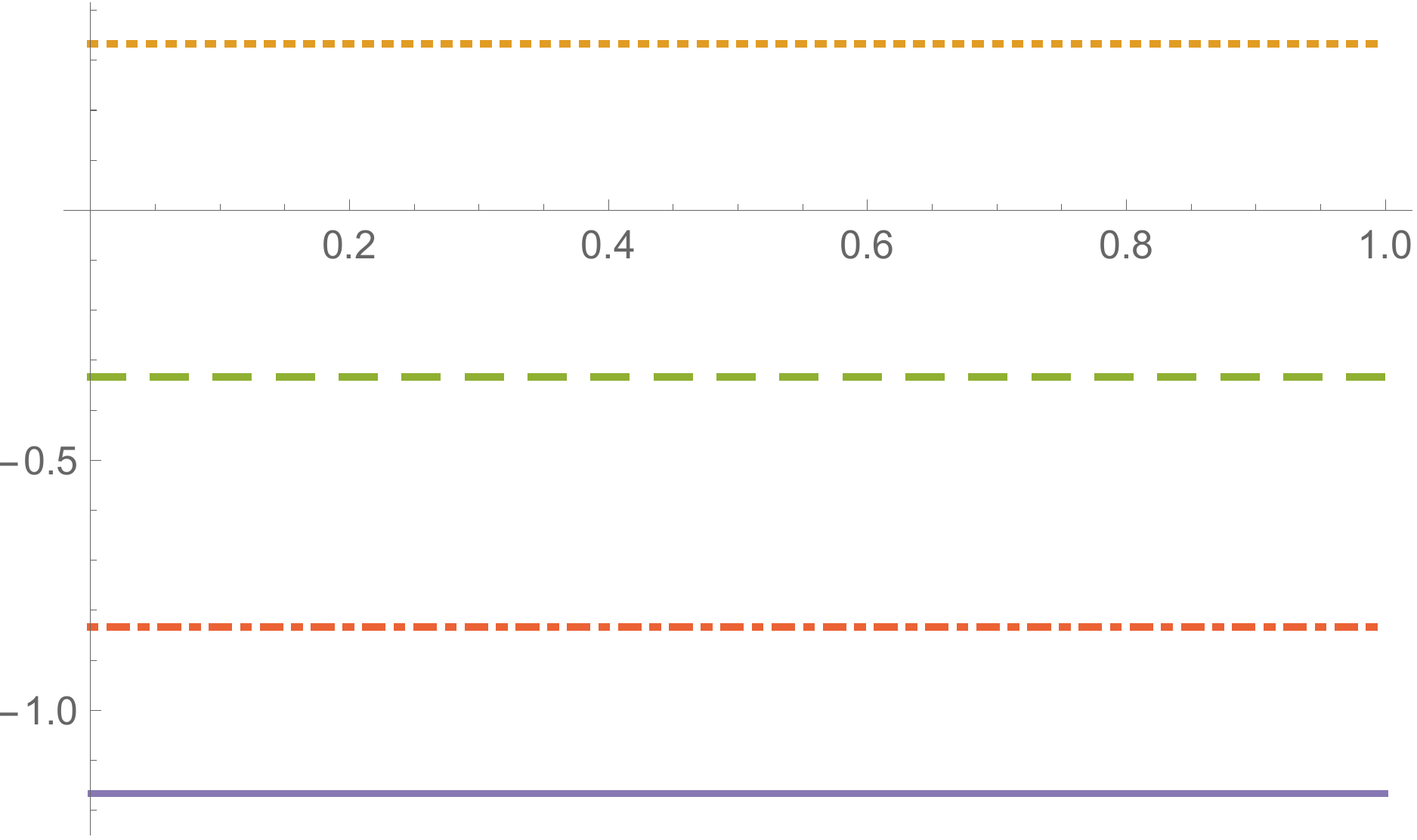}
    \caption{Sound speed versus radius ($r$): Tolman III}
\end{figure}

\newpage
\begin{figure}
    \centering
  \includegraphics[width=0.3\textwidth, height=0.23\textheight]{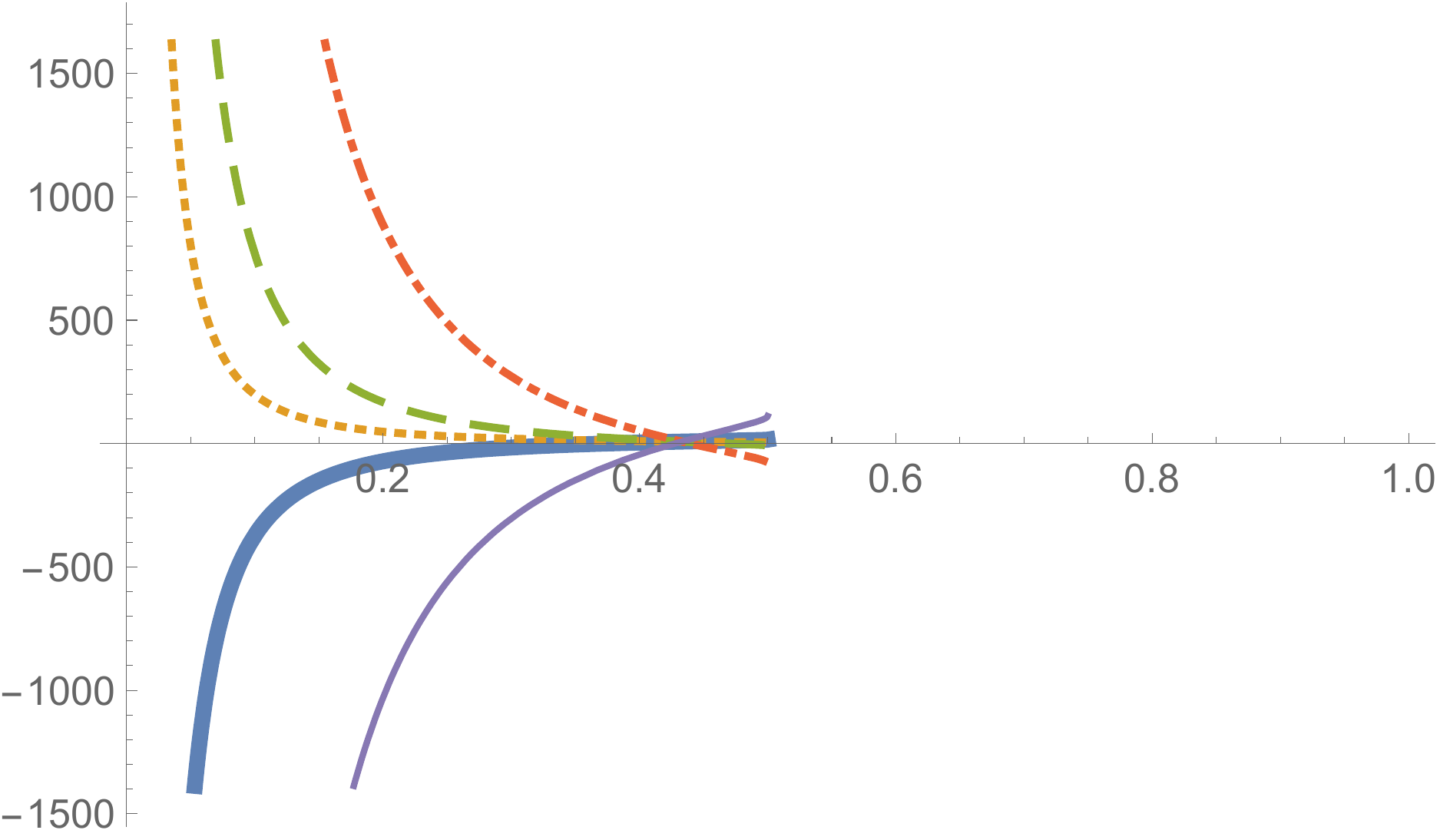}
    \caption{Weak energy condition versus radius ($r$): Tolman III}
\end{figure}

\newpage
\begin{figure}
    \centering
   \includegraphics[width=0.3\textwidth, height=0.23\textheight]{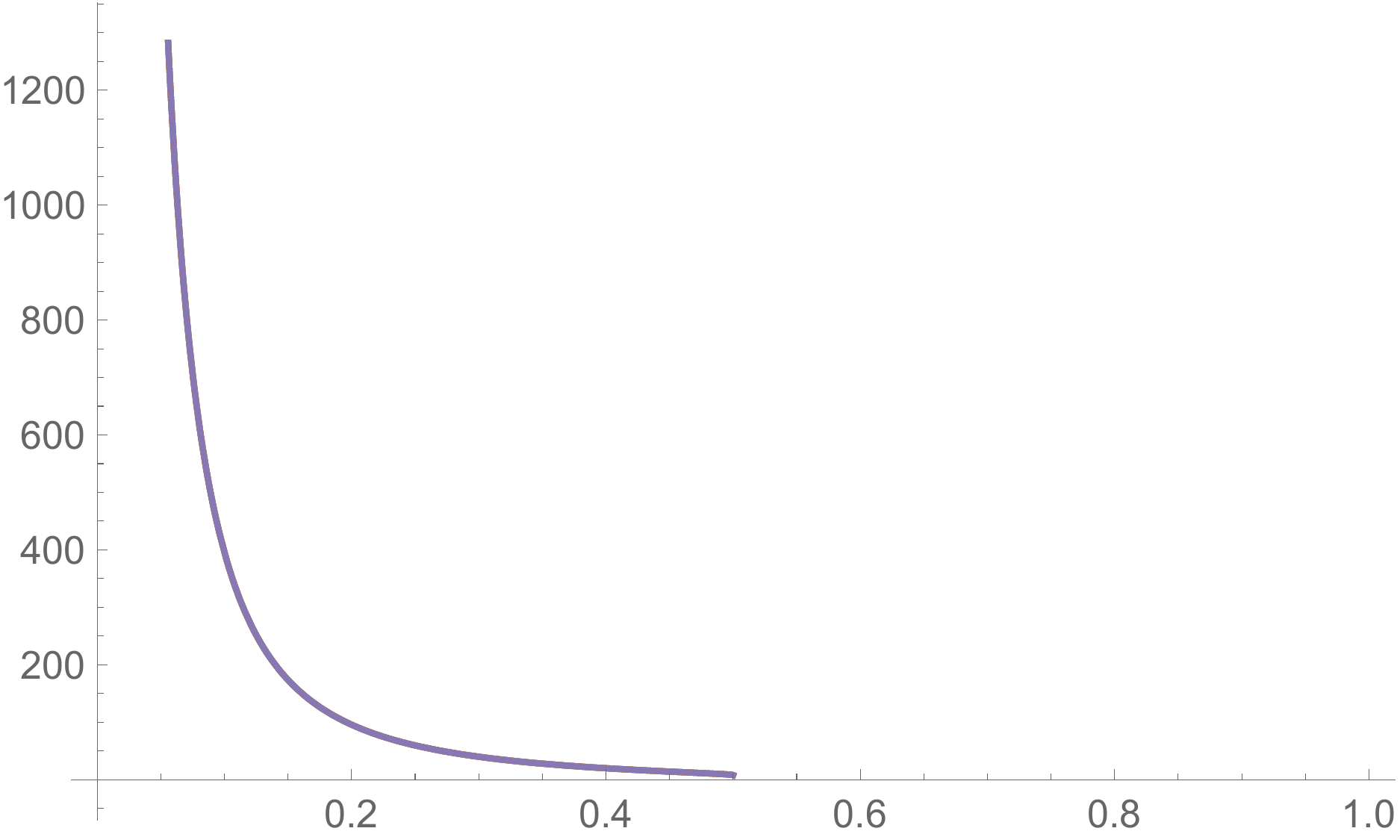}
    \caption{Strong energy condition versus radius ($r$): Tolman III}
\end{figure}

\newpage
\begin{figure}
    \centering
\includegraphics[width=0.3\textwidth, height=0.23\textheight]{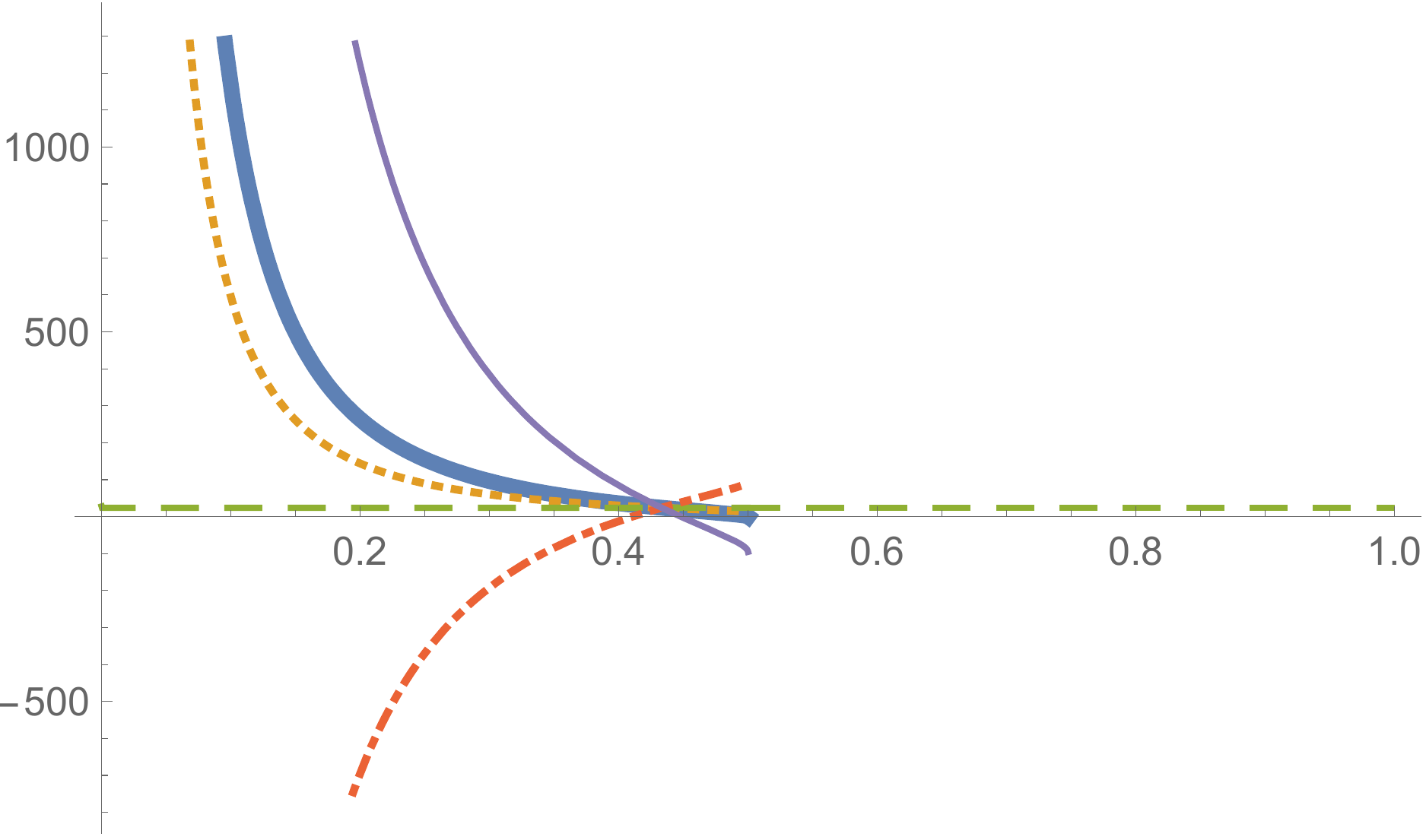}
    \caption{Dominant energy condition  versus radius ($r$): Tolman III}
\end{figure}

\newpage
\begin{figure}
    \centering
   \includegraphics[width=0.3\textwidth, height=0.23\textheight]{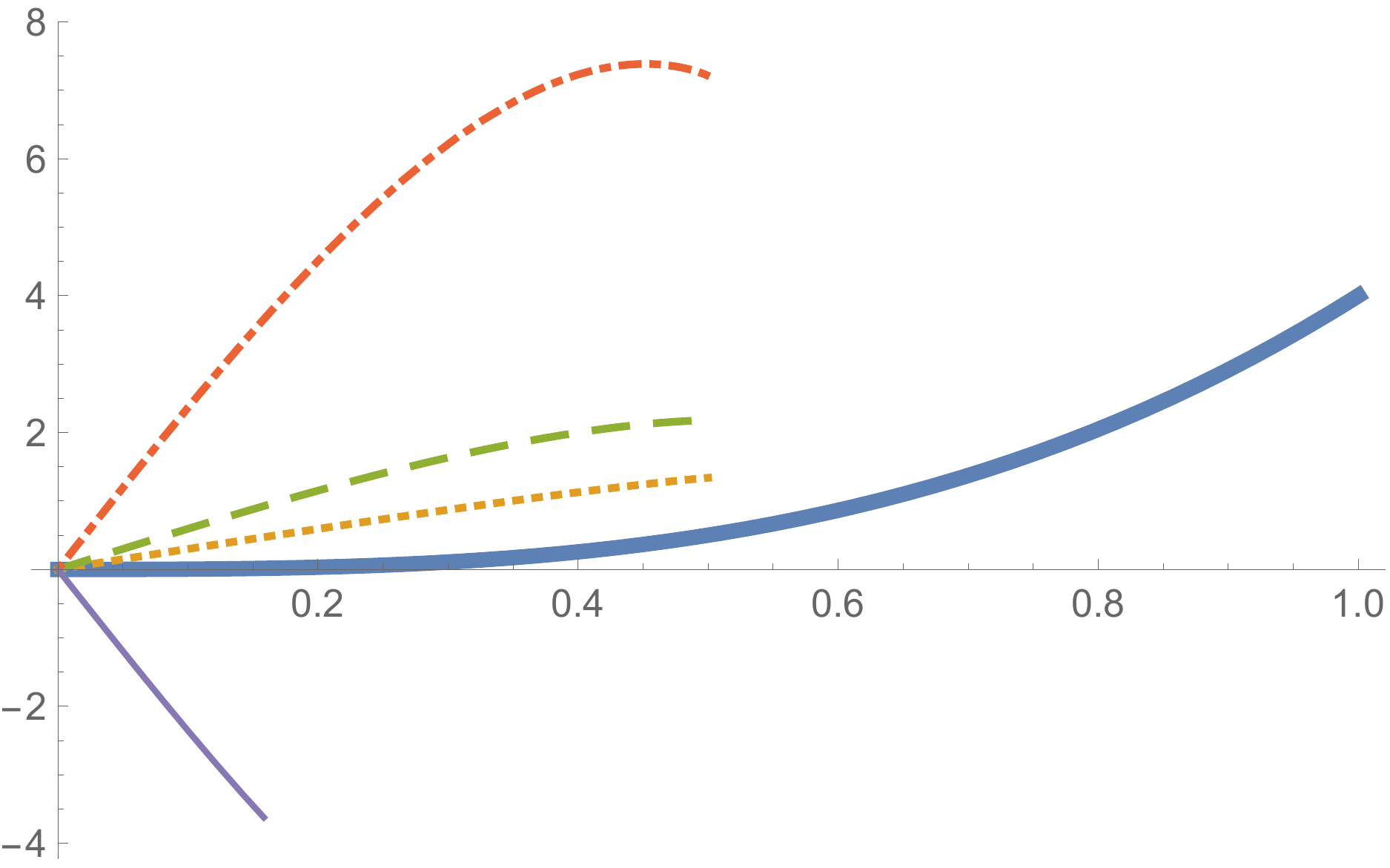}
    \caption{Mass versus radius ($r$): Tolman III}
\end{figure}

Fig. 1 depicts the familiar constant energy density of the Schwarzschild interior solution. However, the Rastall model displays distinct behaviours which exhibit monotonically decreasing functions for positive $\alpha$. The pressure is shown in Fig. 2. We observe that the Einstein and two Rastall cases show similar behaviour for $\alpha=0.25$ and $-2$. Note that the curves reach a zero pressure surface for some radial value. The sound speed is causal for the Rastall case $\alpha = 2$ while it is infinite in the Einstein theory. Figs. 4, 5, 6 show the energy conditions. The Einstein case violates the weak energy condition, while some Rastall cases satisfy all energy conditions. The mass profiles in Fig 7 are suitable in all models except one Rastall case in which the mass is negative. 


\subsection{Tolman IV metric}

The assumption $\frac{e^{\nu}\nu'}{2r}=\text{const.}$
 generated the  metric potentials
\begin{eqnarray}
e^{\lambda}&=&\frac{1+\frac{2 r^2}{A^2}}{\left(1+\frac{r^2}{A^2}\right) \left(1-\frac{r^2}{R^2}\right)}, \nonumber \\
e^{\nu}&=&B^2 \left(1+\frac{r^2}{A^2}\right),
\end{eqnarray}
where $A$, $B$ and $R$ are constants. Within the framework of the Rastall theory the  dynamical quantities
are given by
\begin{eqnarray}
\rho&=&V^{-1}\Big((3-6 \alpha ) A^4+A^2 \left((7-22 \alpha ) r^2+3 R^2\right)\nonumber\\&&+2 r^2 \left((3-12 \alpha ) r^2+(2 \alpha +1) R^2\right)\Big), \label{40a}
\end{eqnarray}
and
\begin{eqnarray}
p&=&V^{-1}\Big((6 \alpha -1) A^4+A^2 \left((22 \alpha -5) r^2+R^2\right)\nonumber\\&&+2 r^2 \left(3 (4 \alpha -1) r^2+(1-2 \alpha ) R^2\right)\Big), \label{40b}
\end{eqnarray}
where we have defined $V=R^2 \left(A^2+2 r^2\right)^2$. The sound speed squared has the form
\begin{equation}
\frac{dp}{d\rho} = \frac{(2 \alpha +1) A^2+2 (1-2 \alpha ) r^2}{(5-2 \alpha ) A^2+2 (2 \alpha +1) r^2}, \label{41}
\end{equation}
while the energy conditions are
\begin{widetext}
\begin{eqnarray}
\rho -p&=&V^{-1}\Big(2 \Big((2-6 \alpha ) A^4+A^2 \left((6-22 \alpha ) r^2+R^2\right)+(6-24 \alpha ) r^4+4 \alpha  r^2 R^2\Big)\Big), \label{42a}\\
\rho +p&=&V^{-1}\Big(2 \left(A^2+r^2\right) \left(A^2+2 R^2\right)\Big),  \label{42b}\\
\rho +3p&=&V^{-1}\Big(2 \Big(6 \alpha  A^4+A^2 \left((22 \alpha -4) r^2+3 R^2\right) +2 r^2 \left(3 (4 \alpha -1) r^2-2 (\alpha -1) R^2\right)\Big)\Big). \label{42c}
\end{eqnarray}
\end{widetext}
The stellar mass varies as the function
\begin{eqnarray}
&&m(r)=V_{1}^{-1}\Big(6 \alpha  A^4 r+A^2 \left(12 \alpha  r R^2-8 (\alpha -1)
r^3\right)\nonumber\\&&-3 \sqrt{2} \alpha  A w +8 r^3 \left((1-4 \alpha ) r^2+(2
\alpha +1) R^2\right)\Big),
\end{eqnarray}
in geometric units and where we have defined $V_{1}=8 R^2 (A^2 + r^2)$ and $w(r) = A^2 + 2r^2
\left(A^2+2 R^2\right) \tan ^{-1}\left(\frac{\sqrt{2} r}{A}\right)$. We have utilised the parameter values $A = B = 1$ and $R = 2$ to generate the plots in Mathematica XI \cite{mathematica}.

\begin{figure}
    \centering
    \includegraphics[width=0.3\textwidth, height=0.23\textheight]{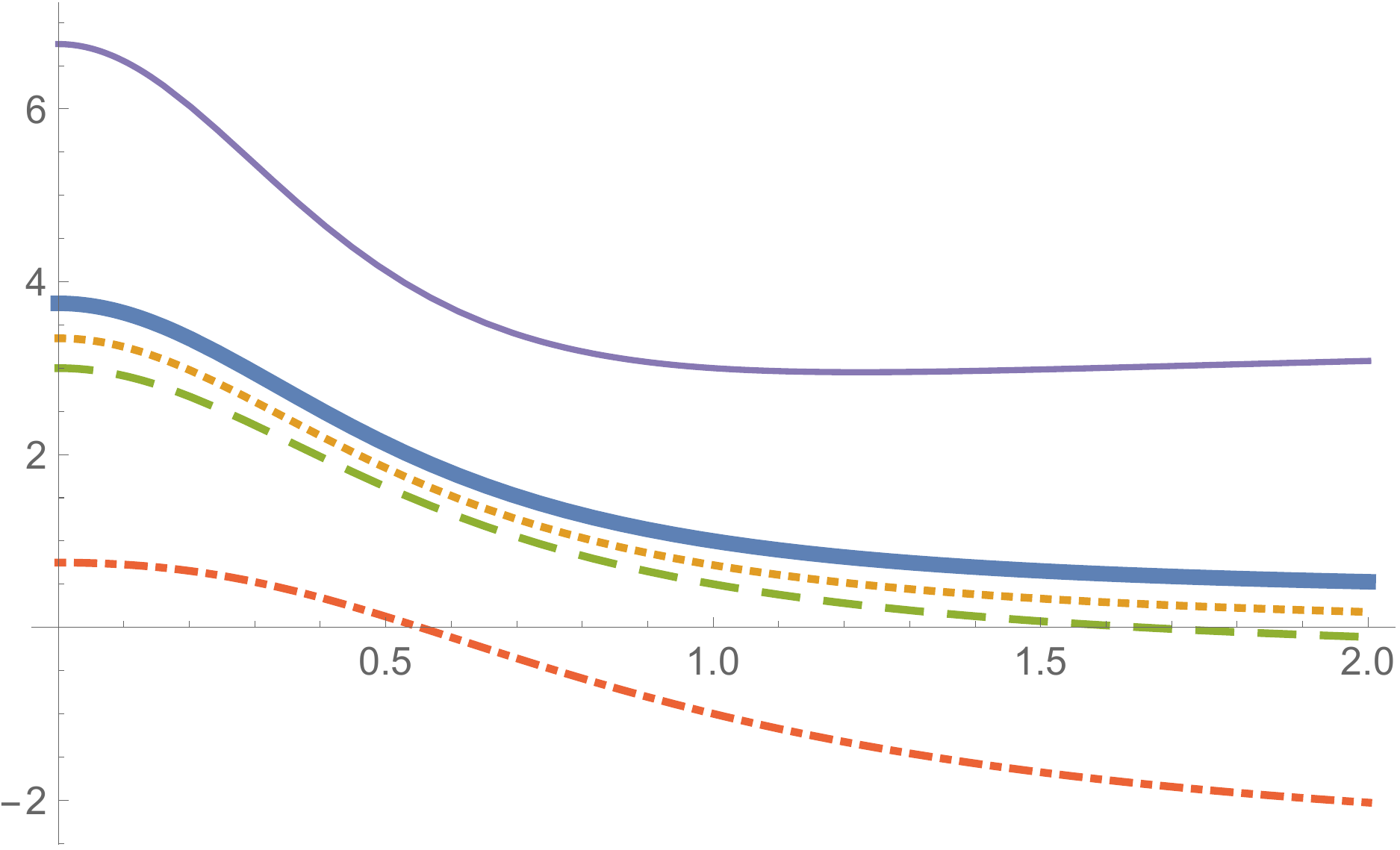}
    \caption{Plot of energy density ($\rho$) versus radius ($r$): Tolman IV}
\end{figure}

\newpage
\begin{figure}
    \centering
 \includegraphics[width=0.3\textwidth, height=0.23\textheight]{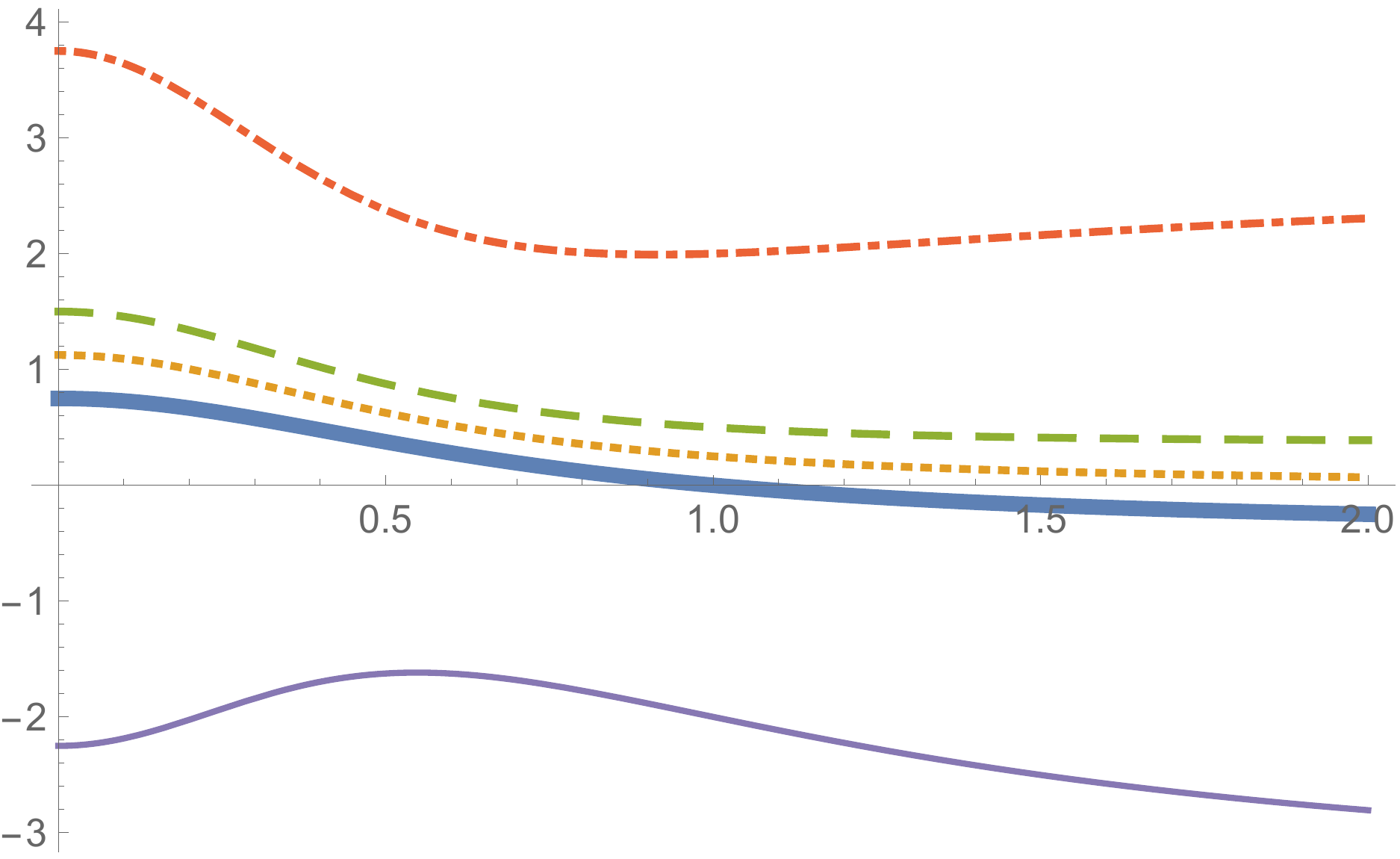}
    \caption{Plot of pressure ($p$) versus radius ($r$): Tolman IV}
\end{figure}

\newpage
\begin{figure}
    \centering
 \includegraphics[width=0.3\textwidth, height=0.23\textheight]{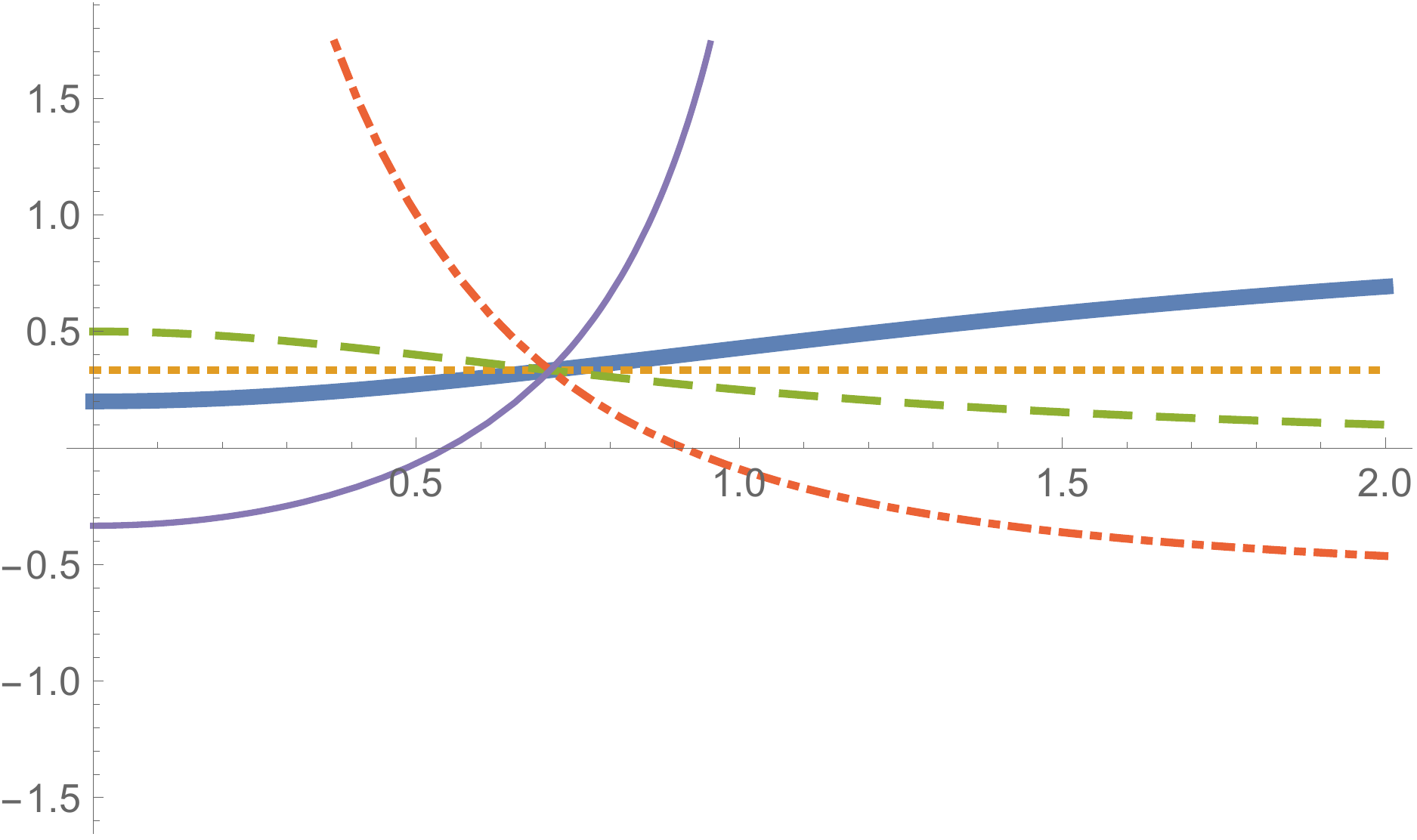}
    \caption{Sound speed versus radius ($r$): Tolman IV}
\end{figure}

\newpage
\begin{figure}
    \centering
  \includegraphics[width=0.3\textwidth, height=0.23\textheight]{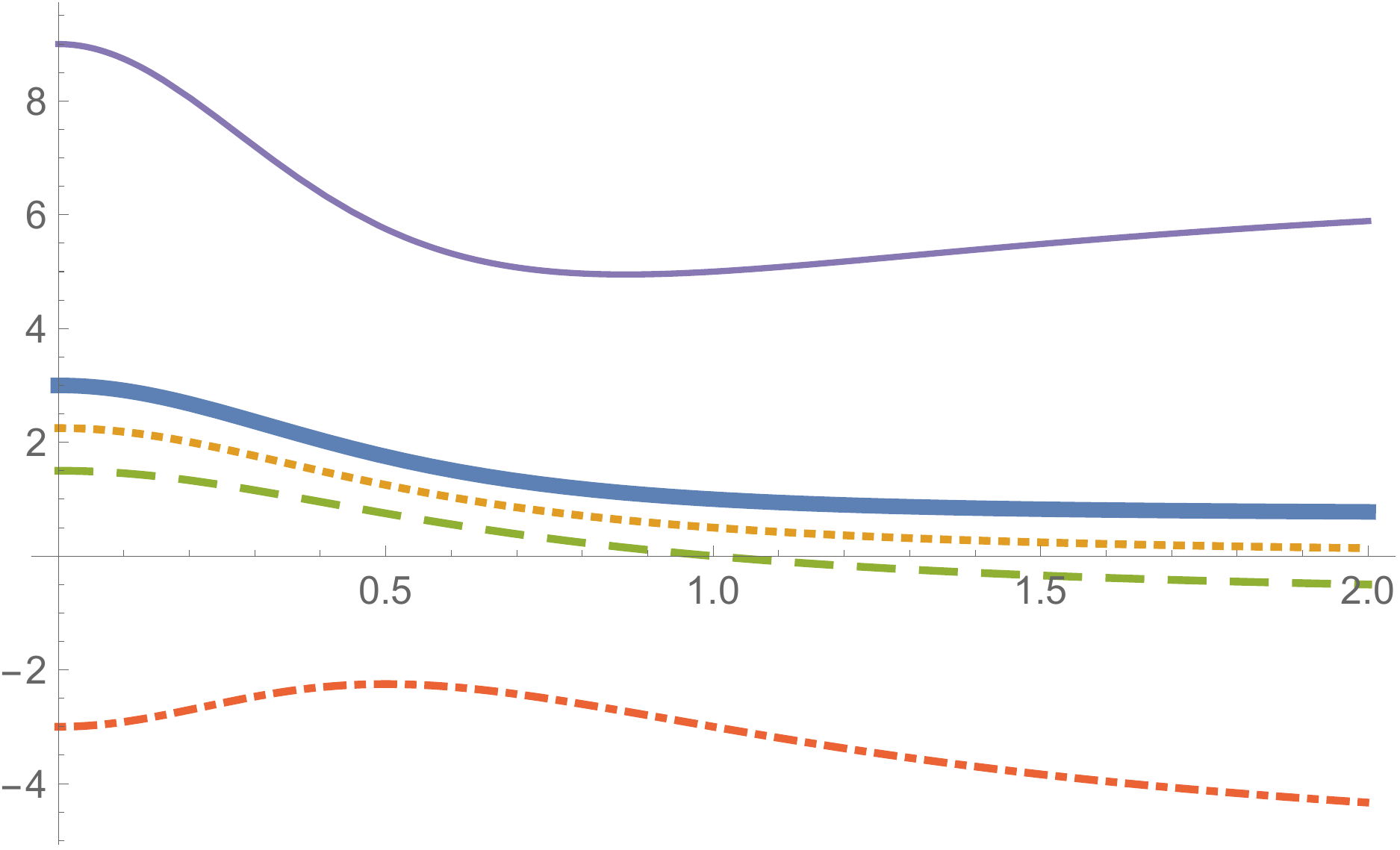}
    \caption{Weak energy condition versus radius ($r$): Tolman IV}
\end{figure}

\newpage
\begin{figure}
    \centering
 \includegraphics[width=0.3\textwidth, height=0.23\textheight]{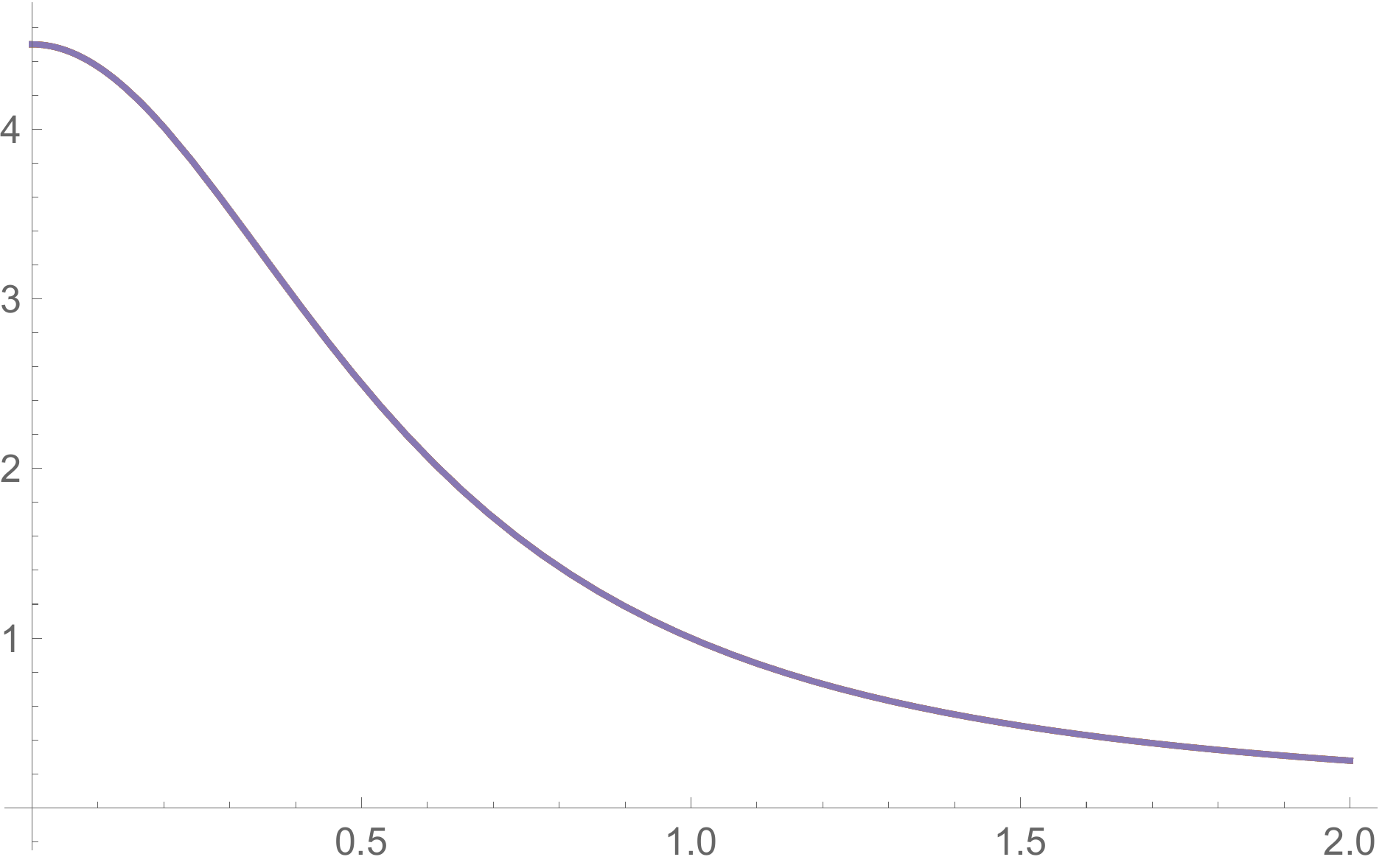}
    \caption{Strong energy condition versus radius ($r$): Tolman IV}
\end{figure}

\newpage
\begin{figure}
    \centering
 \includegraphics[width=0.3\textwidth, height=0.23\textheight]{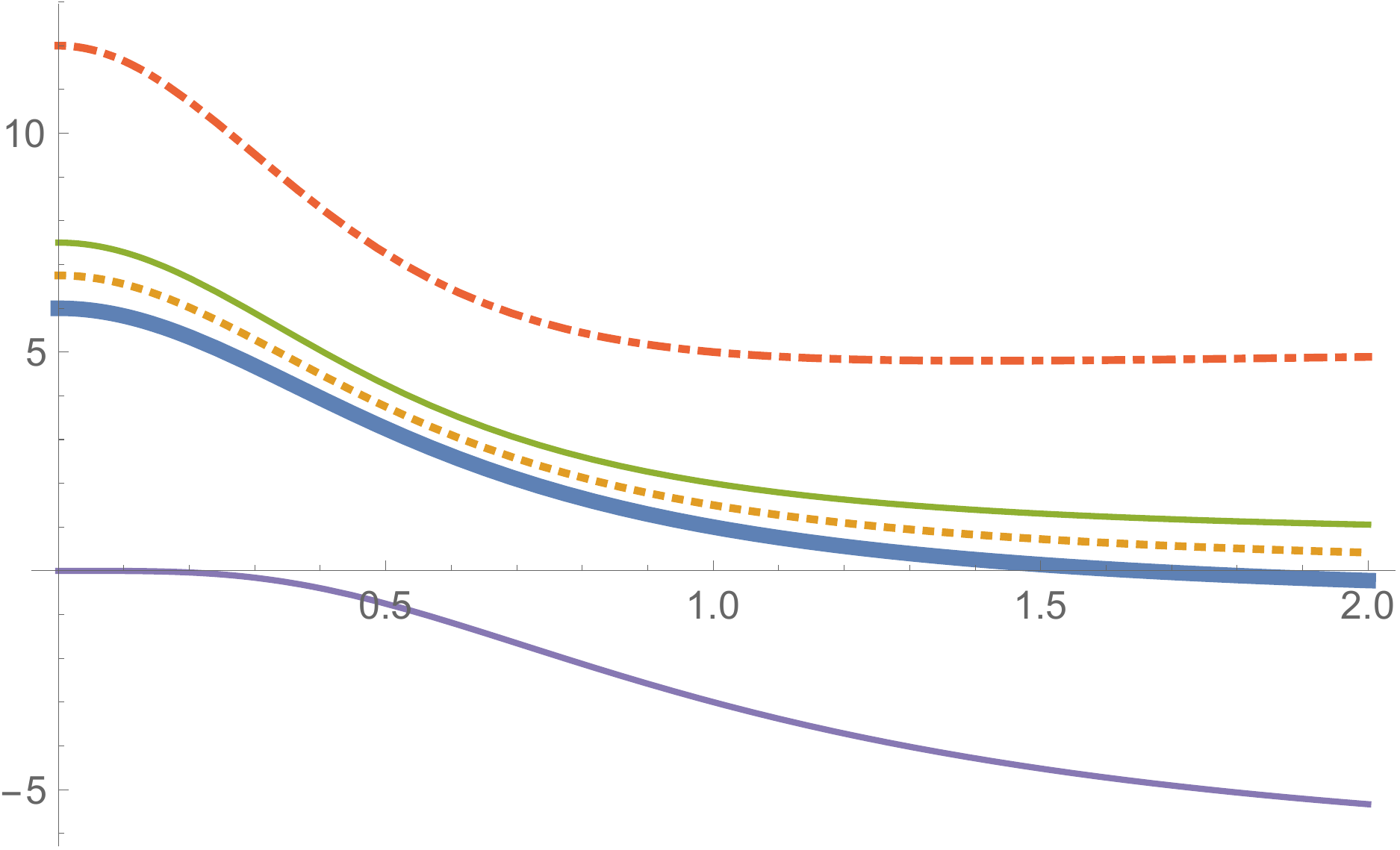}
    \caption{Dominant energy condition  versus radius ($r$): Tolman IV}
\end{figure}

For this case, it is noted that the Einstein model is very well behaved. There exists a pressure free surface at $r = 1$ geometric units and within this bound (Fig 9), the density (Fig 8) and energy expressions (Fig 10, 11, 12) are all positive. The sound speed (Fig 10) is causal having a value between 0 and unity. It is not easy to integrate out the mass function explicitly. It must also be observed that for all Rastall parameters except $\alpha = - 2$, generally pleasing physical behaviour is evident. While some of the Rastall models are superluminal in certain regions where the case $\alpha = 0.25$ and 2 satisfy the causality criterion at all points in the interior. 


\subsection{Tolman V metric}

In this case, Tolman assumes $e^{v} =\text{const.}r^{2n} $ which generates the
potentials of the form:\be e^\lambda=\frac{1+2n-n^2}{1-\left(1+2n-n^2\right)
\left(\frac{r}{R}\right)^{N}} \hspace{0.5cm}  \mbox{and}
\hspace{0.5cm} e^{\nu}=B^2 r^{2n}, \ee
where $n$, $N=\frac{2 \left(1+2n-n^2\right)}{n+1}$, $R$ and $B$ are constants. The dynamical quantities in Rastall gravity take the form
\begin{widetext}
\begin{eqnarray}
\rho&=&\frac{\left(v+1\right) \left(-\frac{2 (\alpha -1) \left(n^2-2 n-1\right) v}{(n+1) \left(v +1\right)}+(4 \alpha -1) \left(1-\frac{-n^2+2 n+1}
{v +1}\right)+6 \alpha  n\right)}{\left(-n^2+2 n+1\right) r^2} ,\label{50a} \\ \nonumber \\
p&=& \frac{\left(v +1\right) \left(\frac{2 \alpha  \left(n^2-2
n-1\right)v}{(n+1) \left(v +1\right)}+(4 \alpha -1)
\left(\frac{-n^2+2 n+1}{v +1}-1\right)+(2-6 \alpha )
n\right)}{\left(-n^2+2 n+1\right) r^2}, \label{50b}
\end{eqnarray}
\end{widetext}
where $v = \left(n^2-2 n-1\right) \left(\frac{r}{R}\right)^{\frac{-2
n^2+4 n+2}{n+1}}$. The sound speed squared is given by
\pagebreak
\begin{widetext}
\begin{eqnarray}
\frac{dp}{d\rho} &=&  W_{2}^{-1} \Bigg(-2 (2 \alpha -1) n^5 w_1 -(2 \alpha +3)
n^4 w_1 +2 \alpha \left(w_2 -3 w_1\right)+2 n^2 \left(-3 \alpha w_2
+ w_2 +w_1\right)  \nonumber\\&&+n \left(w_2 -20
\alpha w_1+4 w_1\right) +n^3 \left(-4 \alpha  \left(w_2 -8
w_1\right)+w_2 -6 w_1\right)+w_1 \Bigg),  \label{51}\\
W_{2} &=& \Big( 2 (2 \alpha +1)
n^5 w_1+(2 \alpha -11) n^4 w_1 -2 \alpha w_2 +6 n^2 \left(\alpha w_2
+ w_1 \right) +n \left(3 w_2  +20 \alpha w_1 -8 w_1 \right) \nonumber\\&&+2 w_2
+n^3 \left(4 \alpha \left(w_2 -8 w_1 \right)-w_2 + 14 w_1\right) +6
\alpha w_1 -3w_1 \Big) ,\label{510}
\end{eqnarray}
\end{widetext}
where we have defined a new variable $W_{2}$ and $w_1 = \left(\frac{r}{R}\right)^{\frac{4
n+2}{n+1}} $ and $w_2 =
\left(\frac{r}{R}\right)^{\frac{2n^2}{n+1}}$.  The expressions
governing the energy conditions assume the forms
\begin{widetext}
\begin{eqnarray}
\rho -p&=& -W_{1}^{-1}\Bigg(2  \Big(4 \alpha  n^4 w_1 +n^2 \left(2 \alpha
\left(w_2-13 w_1\right)+6 w_1\right)+ n \left(-2 \alpha
\left(w_2+13 w_1\right)+w_2
+8 w_1\right)\nonumber\\&&\quad\quad\quad\quad+n^3 \left(4 \alpha w_2 - w_2 + 6 \alpha  w_1 - 4 w_1 \right)-2 (3 \alpha -1) w_1 \Big)\Bigg),\label{52a}\\
\rho + p &=& -\Bigg(2 \Big(2 n^4 w_1 -5 n^3 w_1 +n^2 \left(w_2
-w_1\right) +n \left(w_2 +3 w_1\right) +w_1 \Big)\Bigg) ,  \label{52b}\\
\rho +3p&=& \Bigg(2 \Big(4 (\alpha -1) n^4 w_1 -n \Big(2 \alpha
\left(w_2+13 w_1\right)+w_2 -2 w_1 \Big) +2 n^2 \Big(\alpha
\left(w_2 -13 w_1 \right) \nonumber\\&&\quad\quad-w_2+4 w_1 \Big) +n^3 \left(4 \alpha w_2
-w_2 +6 \alpha w_1 +6 w_1\right) -6 \alpha w_1\Big)\Bigg), \label{52c}
\end{eqnarray}
\end{widetext}
where we have defined $W_{1}=(n+1) w_2
\left(n^2-2 n-1\right) r^2$; while the gravitational mass shows
\begin{eqnarray}
m(r)&&=W^{-1}\Big(n(n-3) r (-2 \alpha +(4 \alpha -1)
n+2)w_2\nonumber\\&&-\left(\left(n^2-2 n-1\right) r (6 \alpha +2 \alpha  n+n-3)
w_1\right)\Big), \label{53}
\end{eqnarray}
where we have defined $W=w_2(-n^2+2 n+1)(n-3)$. The plots have been constructed using the parameter values $n = 2$, $ B = 1$ and $R = 2$. Tolman's choice $n = \frac{1}{2}$ did not generate physically reasonable plots and was accordingly abandoned. 

\begin{figure}
    \centering
   \includegraphics[width=0.3\textwidth, height=0.23\textheight]{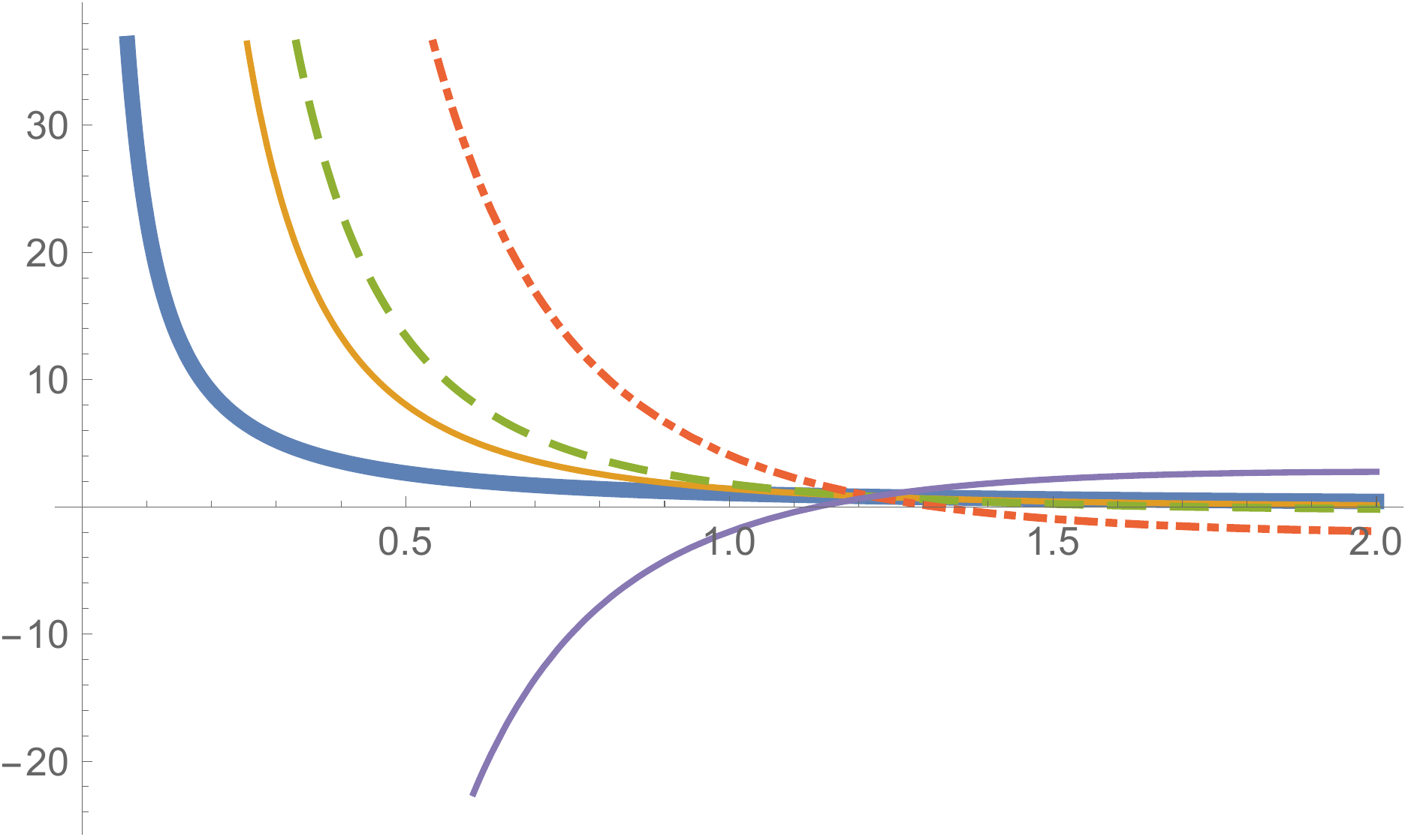}
    \caption{Energy density ($\rho$) versus radius ($r$): Tolman V}
\end{figure}

\newpage
\begin{figure}
    \centering
 \includegraphics[width=0.3\textwidth, height=0.23\textheight]{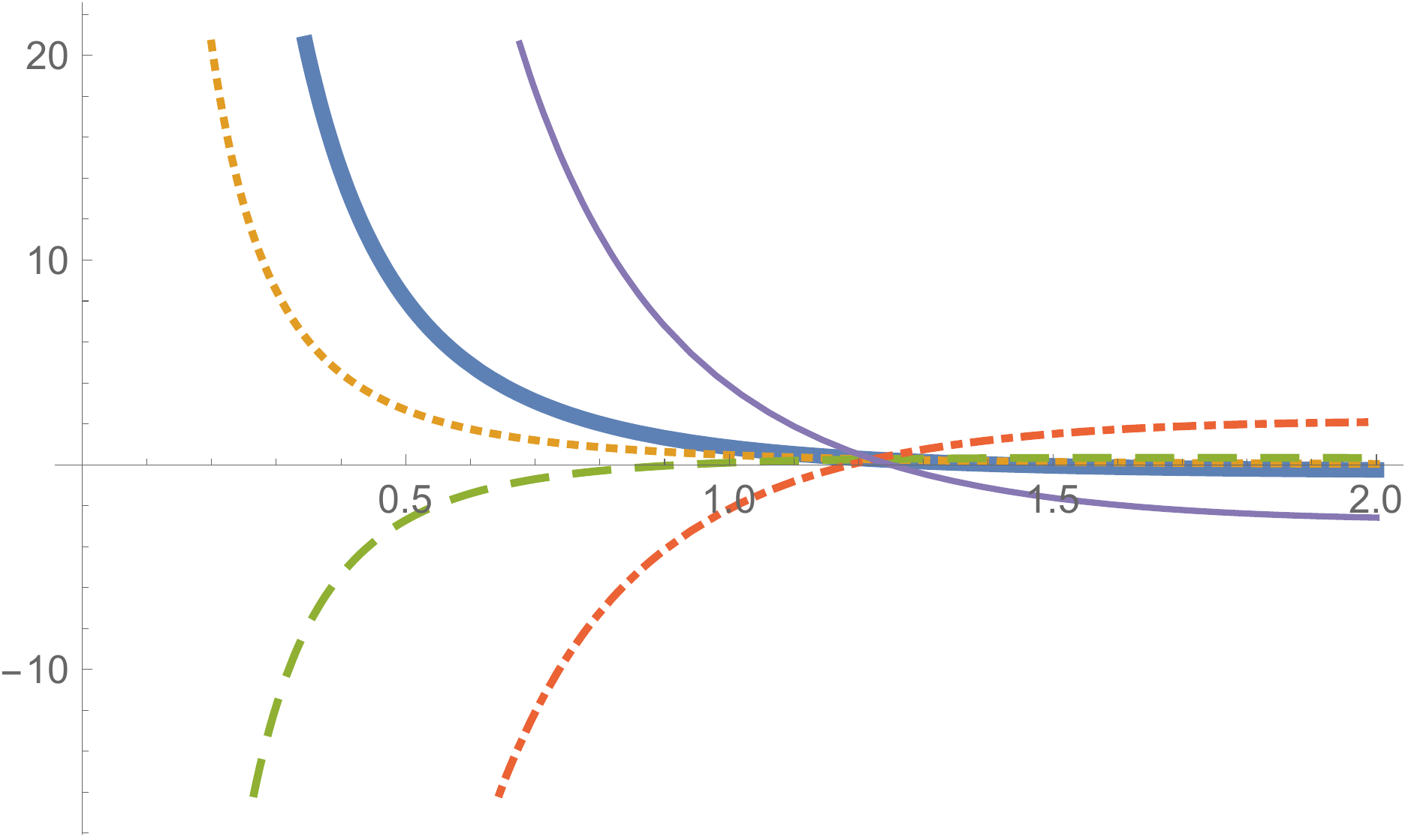}
    \caption{Pressure ($p$) versus radius ($r$): Tolman V}
\end{figure}

\newpage
\begin{figure}
    \centering
 \includegraphics[width=0.3\textwidth, height=0.23\textheight]{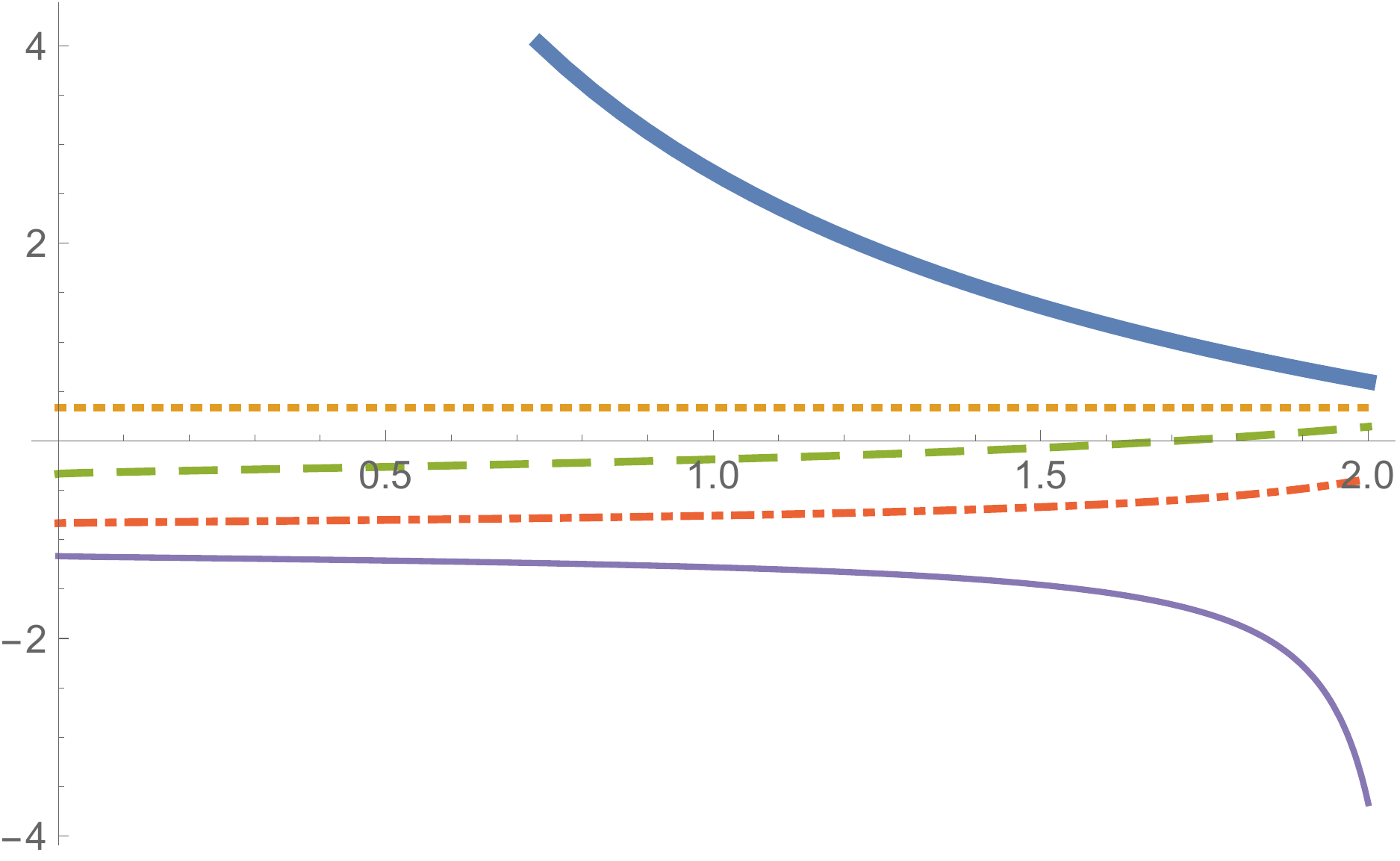}
    \caption{Sound speed versus radius ($r$): Tolman V}
\end{figure}

\newpage
\begin{figure}
    \centering
 \includegraphics[width=0.3\textwidth, height=0.23\textheight]{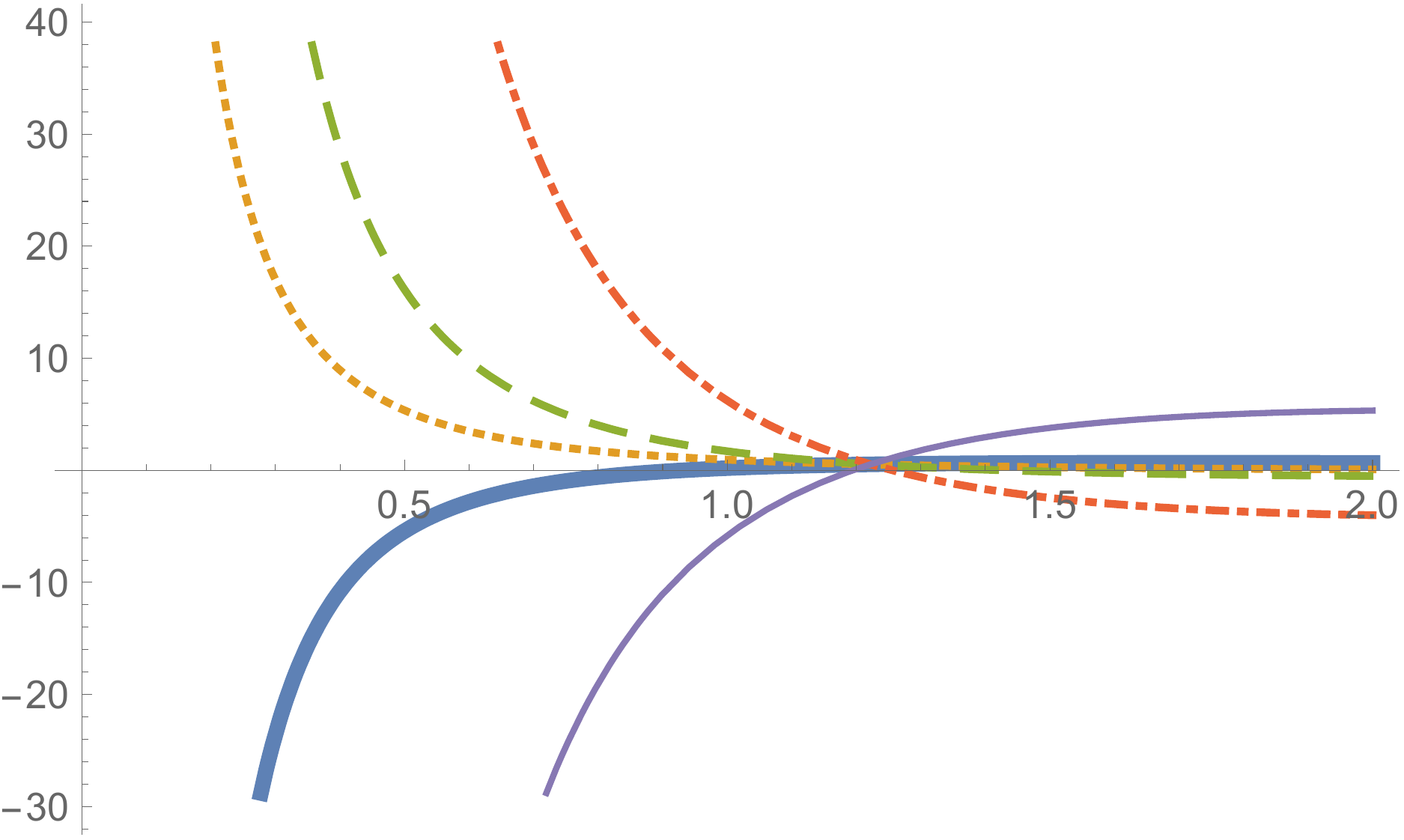}
    \caption{Weak energy condition versus radius ($r$): Tolman V}
\end{figure}

\newpage
\begin{figure}
    \centering
 \includegraphics[width=0.3\textwidth, height=0.23\textheight]{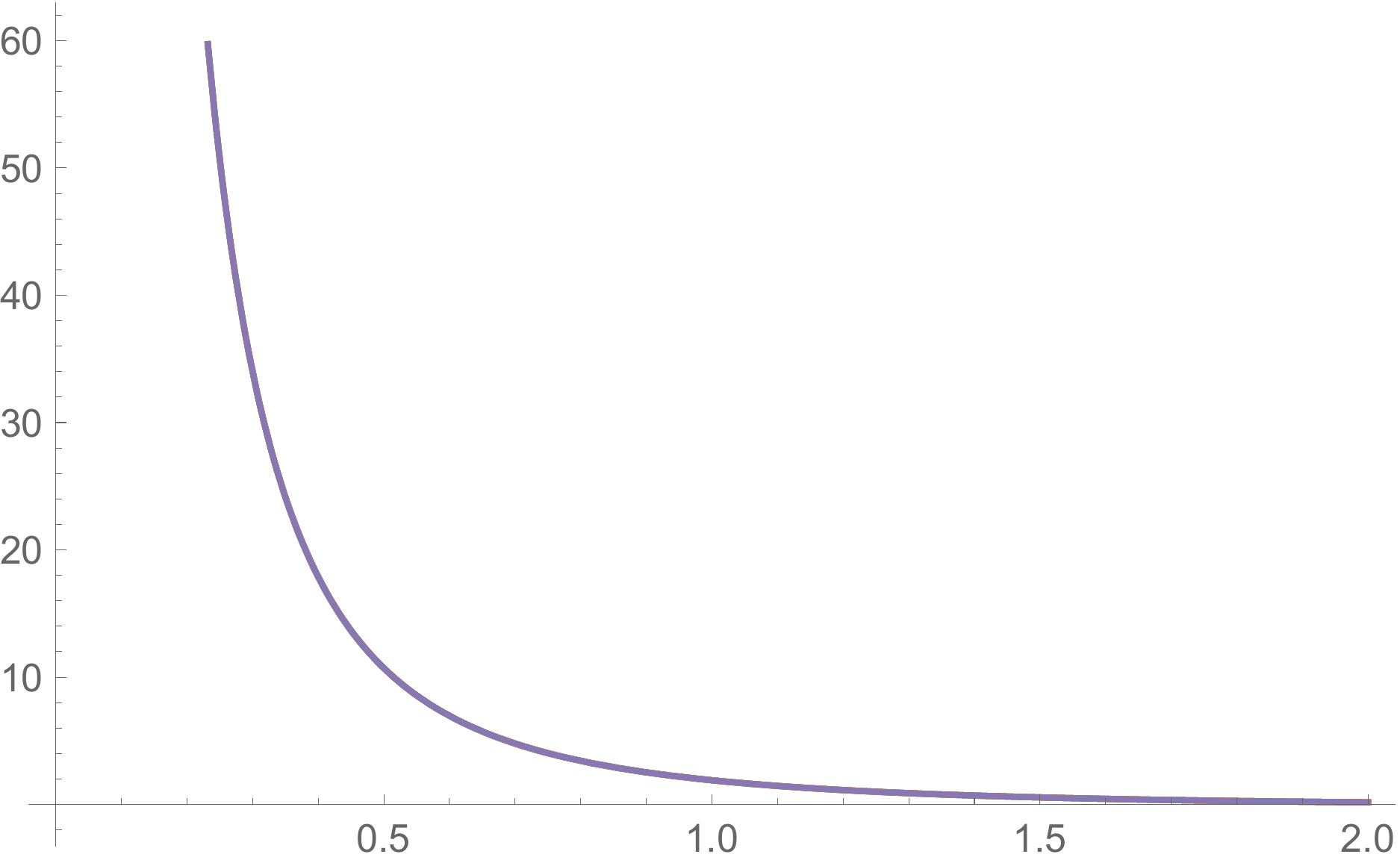}
    \caption{Strong energy condition versus radius ($r$): Tolman V}
\end{figure}

\newpage
\begin{figure}
    \centering
 \includegraphics[width=0.3\textwidth, height=0.23\textheight]{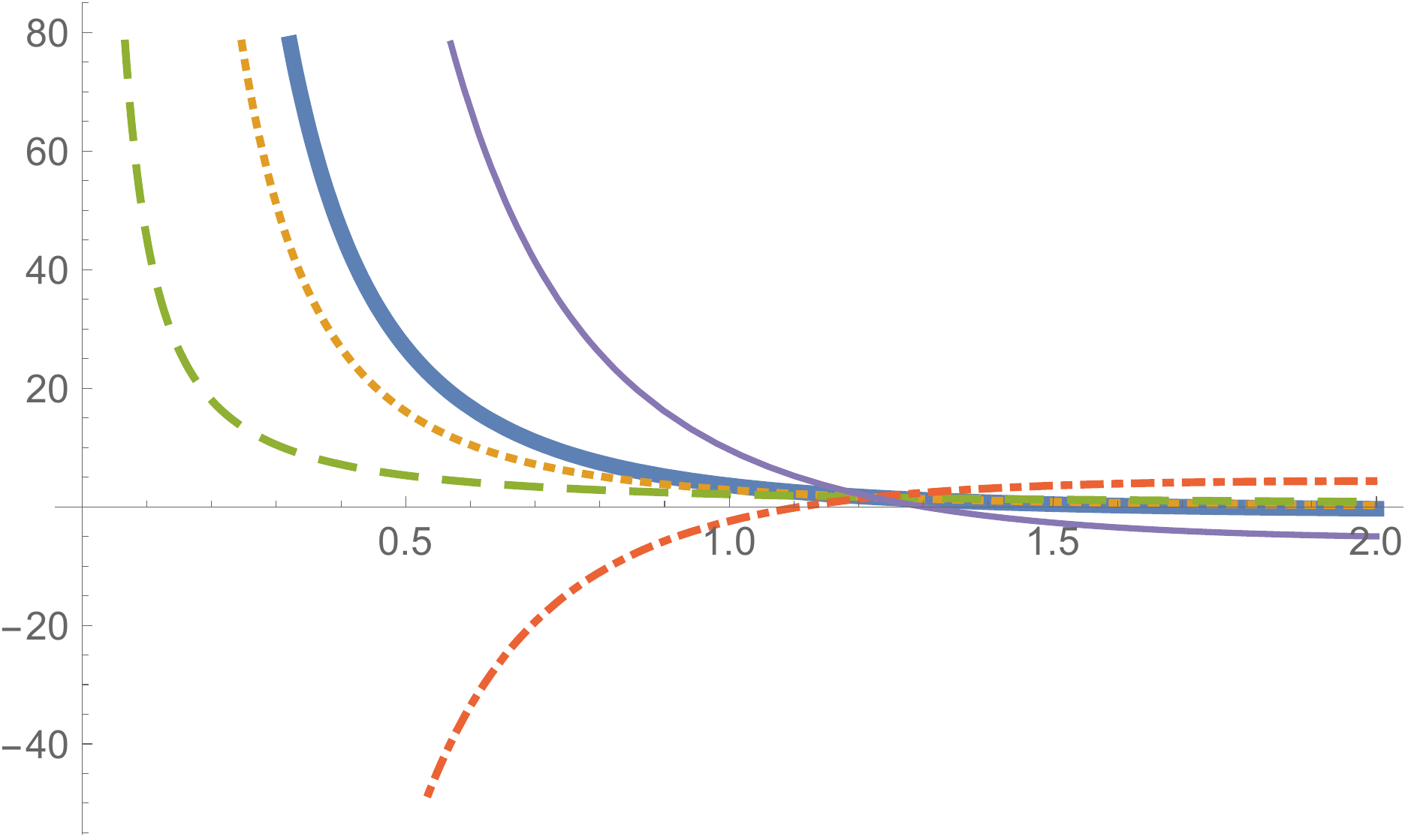}
    \caption{Dominant energy condition  versus radius ($r$): Tolman V}
\end{figure}

For this Tolman model, causality (Fig 16) and weak energy (Fig 17) violation emerge within the distribution in the Einstein case. The energy density (Fig 14) and pressure (Fig 15) are reasonably behaved, with a surface of vanishing pressure. What is interesting in this model is that the Rastall case $\alpha = 0.25$ conforms to all the elementary physical demands including consistent subluminal behaviour as shown in Fig 16. All the energy conditions are also satisfied (Fig 17, 18, 19). This exhibits a case where the Rastall sphere bears a greater resemblance to physical reality similar to that the Einstein sphere. This case is noteworthy in demonstrating the deviation of Rastall theory from general relativity in light of the physical behaviour evident. Clear the Rastall parameter is a useful mathematical handle to correct the shortcomings in the associated general relativity version and this lends credence to our concerns about the claim of Visser that both theories are equivalent. The qualitative evidence in this case does not support that conclusion from a physical perspective. 


\subsection{Tolman VI metric}

The prescription $
    e^{-\lambda}=\text{const.}=\frac{1}{2-n^2}$, $n$ a constant,
 produces the temporal potential $ e^{\nu} = \left(A r^{1-n}-B
r^{n+1}\right)^2 $ where $A$ and $B$ are integration constants.

In the context of Rastall theory, the dynamical quantities assume
the forms
\begin{eqnarray}
\rho=&&\frac{1}{\left(n^2-2\right) r^2\left(A-B r^{2 n}\right)}\times\nonumber\\&& \Big(A \Big(-2 \alpha +(1-4 \alpha ) n^2+6 \alpha  n-1\Big)\nonumber\\&&+B (n+1) (2 \alpha +(4 \alpha -1) n+1) r^{2 n}\Big), \label{60a}
\end{eqnarray}
and
\begin{eqnarray}
p &=&\frac{A (n-1) (\beta_1 + 1) -B (n+1) (\beta_2 - 1) r^{2
n}}{\left(n^2-2\right) r^2 \left(A-B r^{2 n}\right)}, \label{60b}
\end{eqnarray}
where we have put $\beta_1 = (-2 \alpha +(4 \alpha -1) n)$  and
$\beta_2 = (2 \alpha +(4 \alpha -1) n)   $.  The sound speed
parameter is given by
\begin{equation}
\frac{dp}{d\rho}= -\frac{Z+B^2 (n+1) (\beta_2 -1) r^{4 n}}{Z+B^2 (n+1) (\beta_2 +1) r^{4 n}} , \label{61}  
\end{equation}
where we have defined
\begin{eqnarray}
Z=&&A^2 (n-1) (\beta_1 + 1) \nonumber\\&&+2 (2 \alpha -1) A
B \left(n^2-1\right) r^{2 n}.
\end{eqnarray}
The energy conditions may be studied with the help of the 
expressions 
\begin{eqnarray}
&&\rho - p=-\frac{2 \left(A (n-1) \beta_1-B (n+1) \beta_2 r^{2 n}\right)}{\left(n^2-2\right) r^2 \left(A-B r^{2 n}\right)} \label{62a} \\ \nonumber \\
&&\rho + p= \frac{2 \left(A (n-1)+B (n+1) r^{2 n}\right)}{\left(n^2-2\right) r^2 \left(A-B r^{2 n}\right)} \label{62b} \\ \nonumber \\
&&\rho +3p=\frac{2 \left(A (n-1) (\beta_1 +2)-B (n+1) (\beta_2 -2)
r^{2 n}\right)}{\left(n^2-2\right) r^2 \left(A-B r^{2 n}\right)} 
\label{62c}
\end{eqnarray}
while the mass behaviour  given by
 \begin{eqnarray}
M = &&\frac{1}{n^2-2}\Bigg(r \Big(12 \alpha  n \,
_2F_1\left(1,\frac{1}{2 n};1+\frac{1}{2 n};\frac{B r^{2
n}}{A}\right)\nonumber\\&&\quad\quad\quad\quad-(n+1) (2 \alpha +(4 \alpha -1) n+1)\Big)\Bigg),
\label{63}
 \end{eqnarray}
which is formulated in terms of the hypergeometric function
$_2F_1$. For special cases of $n$ elementary functions result.

In order to plot the various dynamical quantities we have used the more general form of the Tolman ansatz that is $e^{\lambda} = K$  for some constant $K$. The selected parameter values were $A = B = 1$ and $K = 1.5$. 

\begin{figure}
    \centering
  \includegraphics[width=0.3\textwidth, height=0.23\textheight]{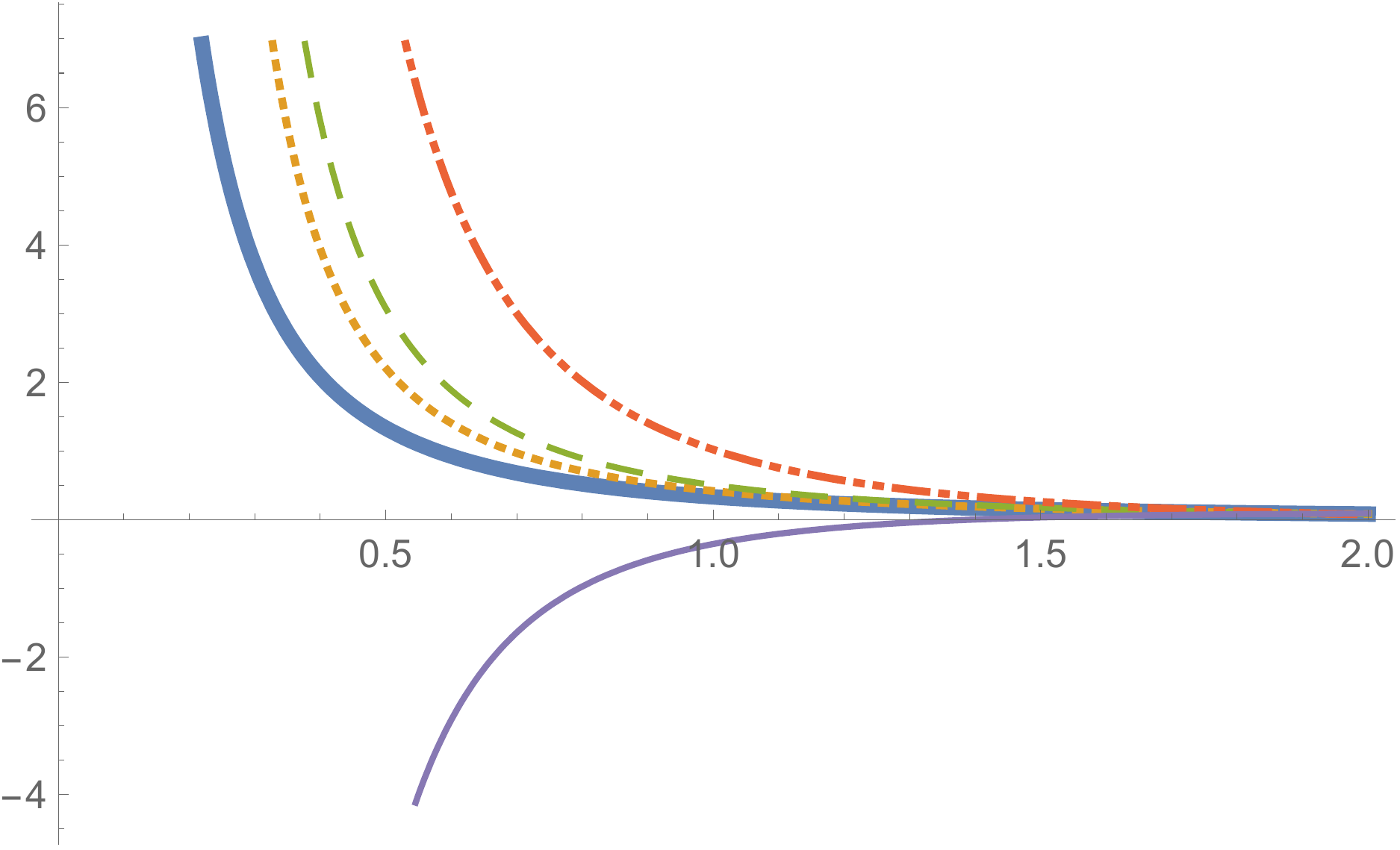}
    \caption{Energy density ($\rho$) versus radius ($r$): Tolman VI}
\end{figure}

\begin{figure}
    \centering
 \includegraphics[width=0.3\textwidth, height=0.23\textheight]{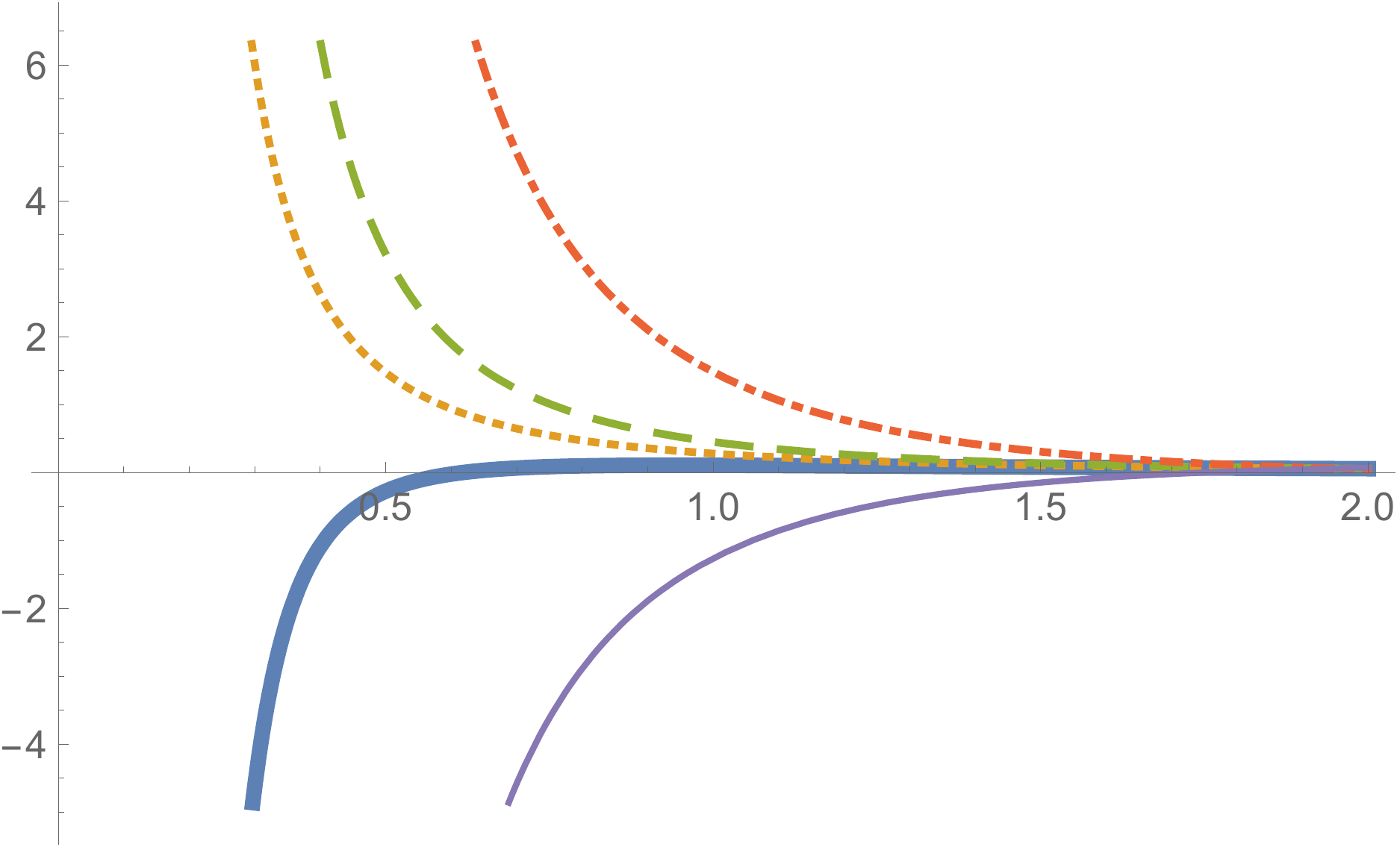}
    \caption{Pressure ($p$) versus radius ($r$): Tolman VI}
\end{figure}

\begin{figure}
    \centering
 \includegraphics[width=0.3\textwidth, height=0.23\textheight]{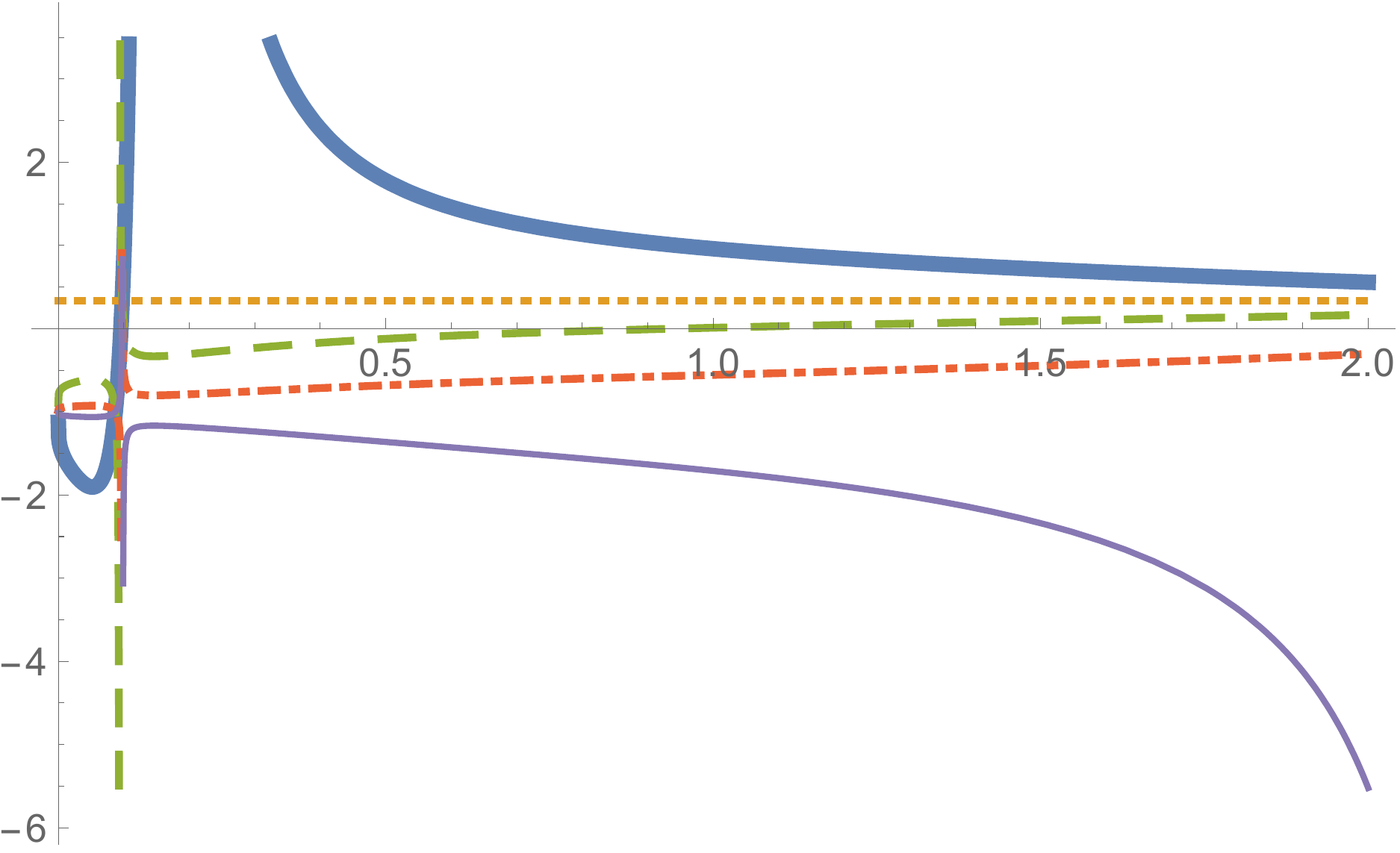}
    \caption{Sound speed versus radius ($r$): Tolman VI}
\end{figure}

\begin{figure}
    \centering
  \includegraphics[width=0.3\textwidth, height=0.23\textheight]{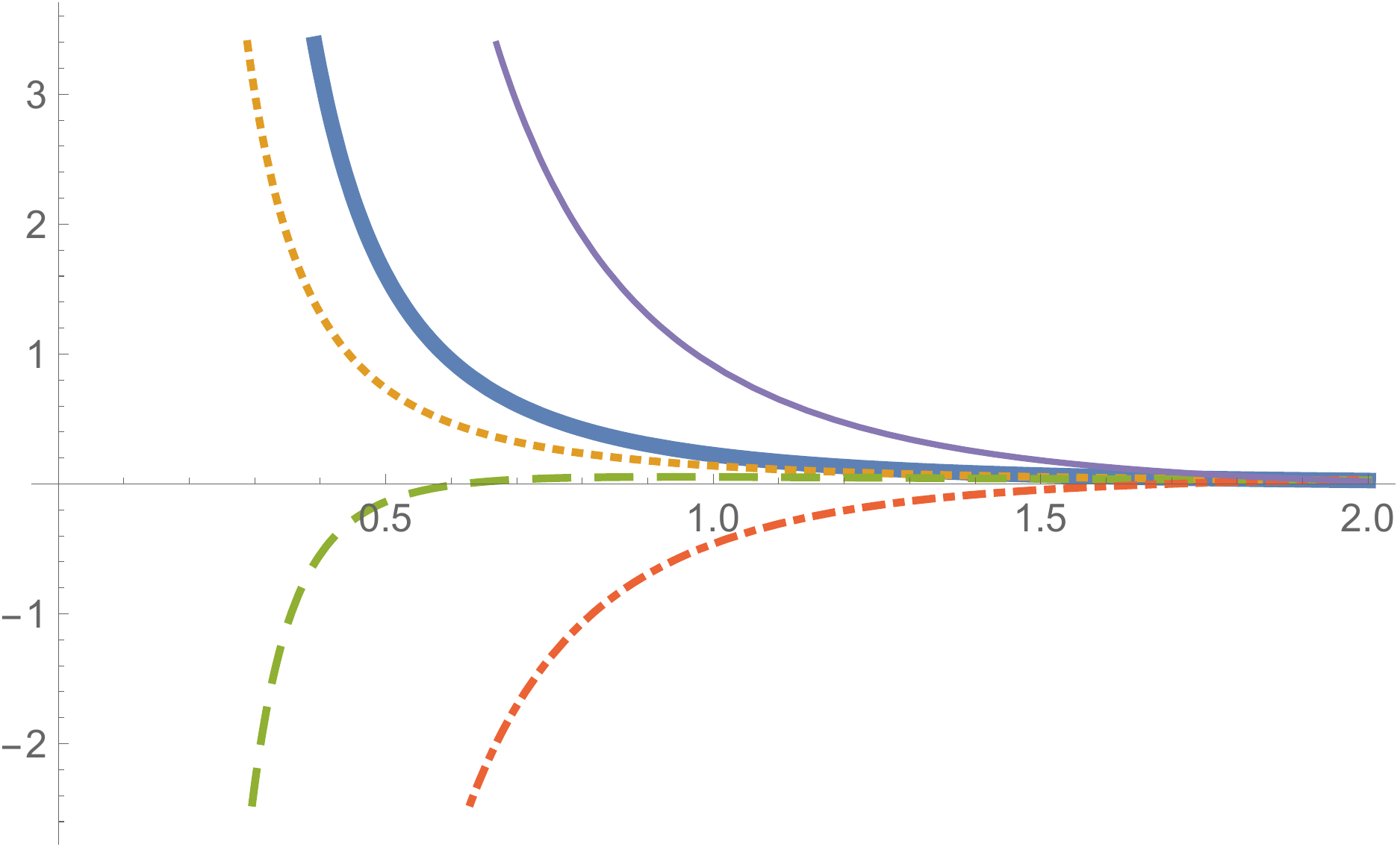}
    \caption{Weak energy condition versus radius ($r$): Tolman VI}
\end{figure}

\begin{figure}
    \centering
 \includegraphics[width=0.3\textwidth, height=0.23\textheight]{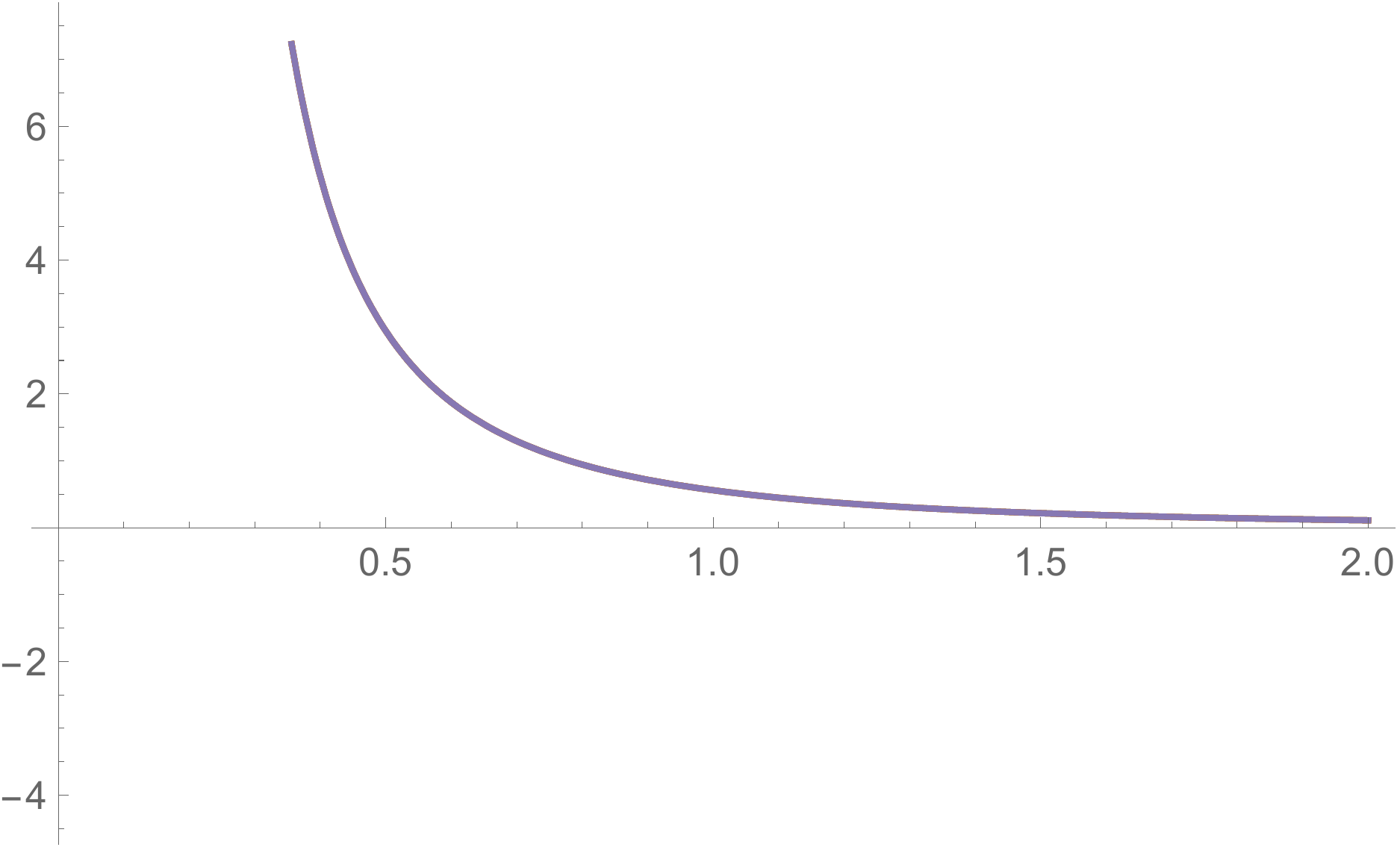}
    \caption{Strong energy condition versus radius ($r$): Tolman VI}
\end{figure}

\begin{figure}
    \centering
 \includegraphics[width=0.3\textwidth, height=0.23\textheight]{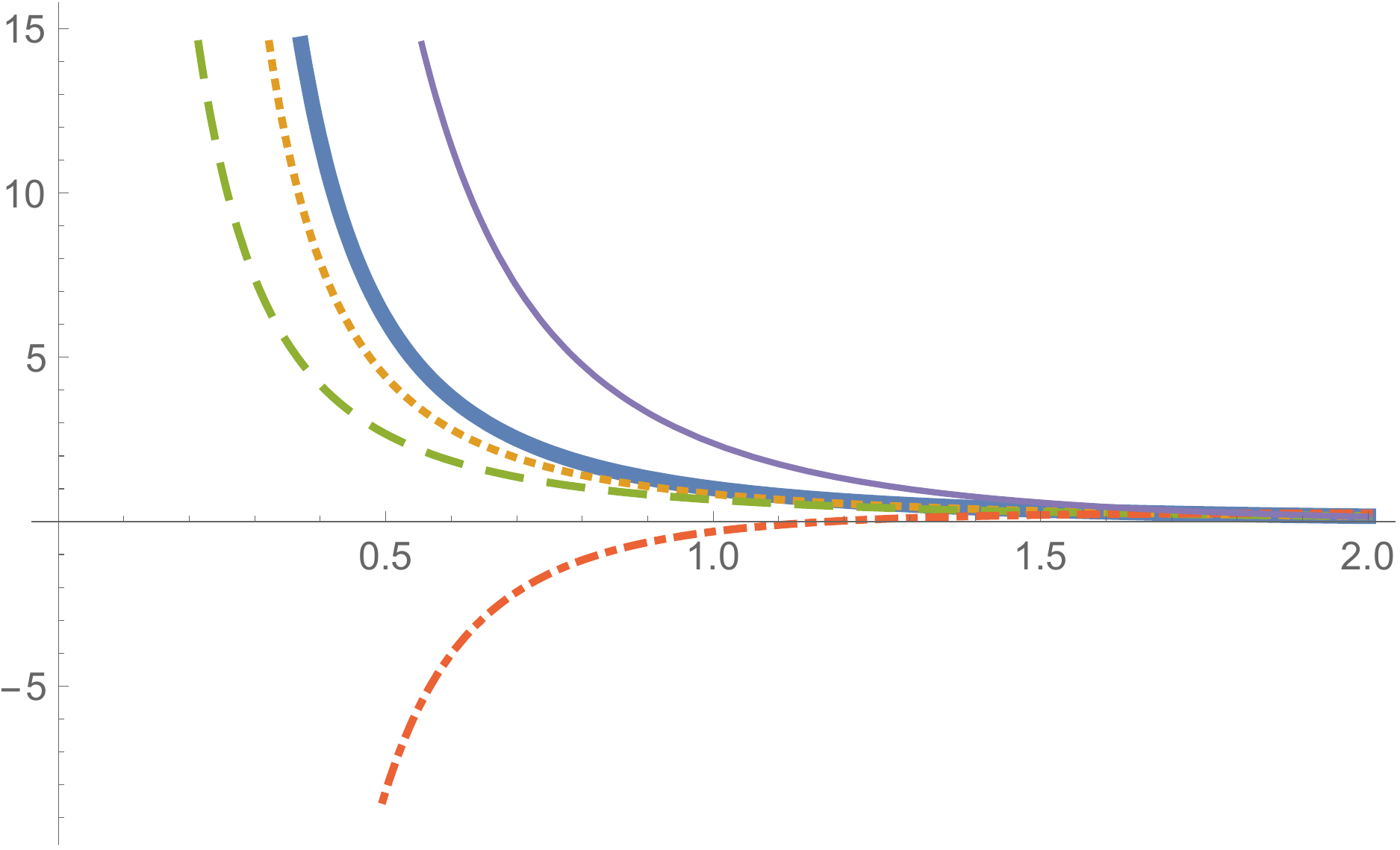}
    \caption{Dominant energy condition  versus radius ($r$): Tolman VI}
\end{figure}

\begin{figure}
    \centering
 \includegraphics[width=0.3\textwidth, height=0.23\textheight]{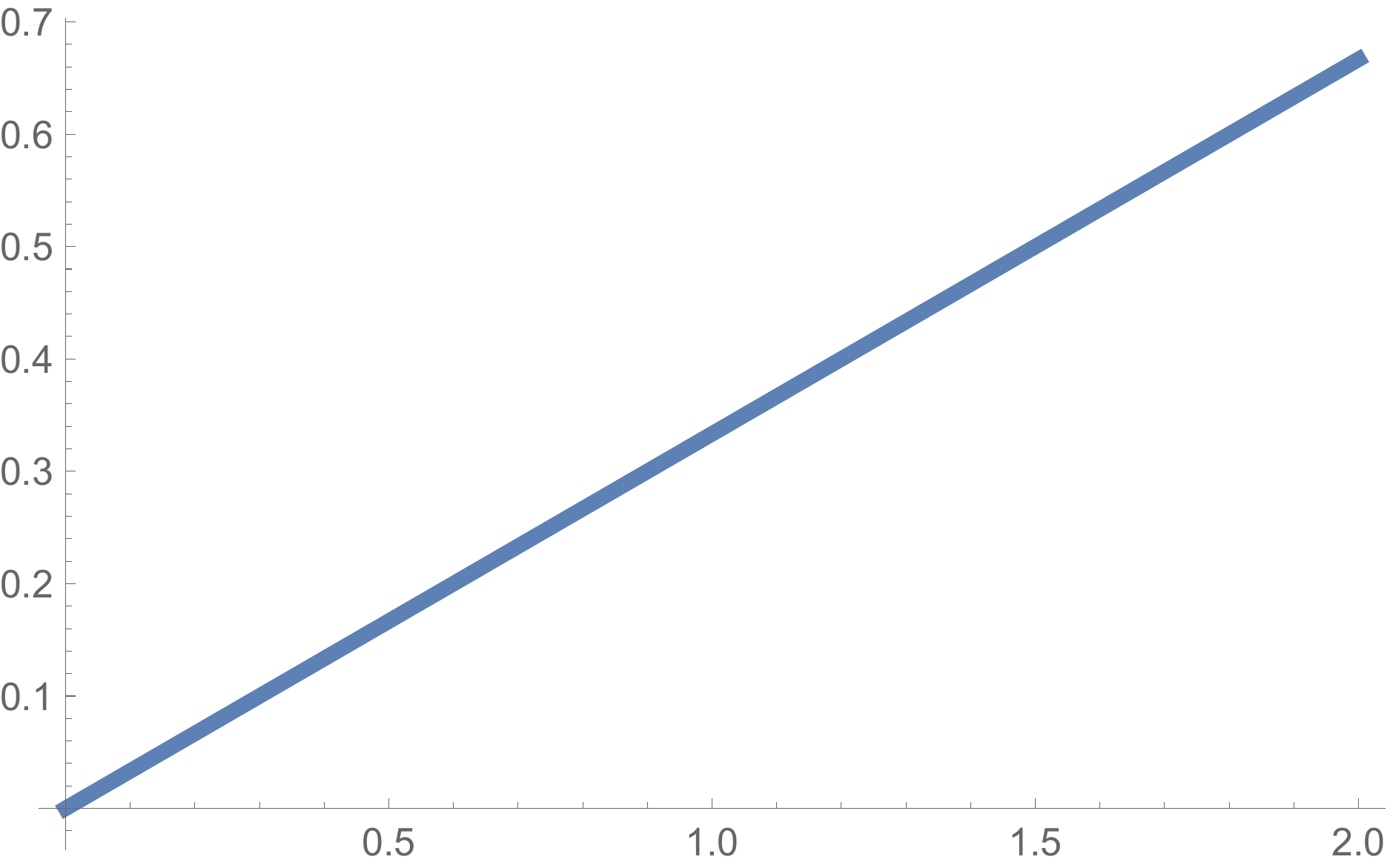}
    \caption{Dominant energy condition  versus radius ($r$): Tolman VI}
\end{figure}

We find that in the Einstein case the density (Fig 20) is positive, while the pressure (Fig 21) is negative for the same parameter values. This is not palatable when modeling stars. Moreover, the sound speed (Fig 22) demonstrates the existence of asymptotes - again these singularities are not expected in regular distributions. In contrast  the Rastall case, $\alpha = 0.25$ displays satisfactory physical behavior. Density and pressure are positive, sound speed obeys  causality and the energy conditions (Fig 23, 24, 25) are satisfied. It is possible to obtain the mass profile (Fig 26) explicitly only for the Einstein case and evidently the mass varies linearly with increasing radius.  Here is another example where a Rastall fluid sphere is well behaved in comparison with its Einstein counterpart. This case also serves as empirical evidence against the Visser claim that the general relativistic models with energy conservation are equivalent physically to the Rastall models which violate energy conservation although they satisfy the basic energy conditions.


\subsection{Extended  Tolman VII metric}

Tolman's assumed spatial potential $
e^{-\lambda}= 1-\frac{r^2}{R^2} + \frac{4r^4}{A^4} $ 
yields the metric potential
\[
e^\nu = B^2\left[\sin \left( \log
\left(\frac{e^{-\frac{\lambda}{2}} + 2r^{2}/A^{2}-A^{2}/4R^{2}}{C}
\right)^{\frac{1}{2}} \right) \right]^2.
\]

Despite the polynomial assumption  that Tolman made, the
solution became lengthy and intractable. For this reason, it was not
included in his paper. To find the expanded solution, we substitute 
the form for $\lambda$ into the isotropy equation (\ref{6c}) to give
\begin{equation}
    e^{\nu}= H\left(K\cos{f}+\sin{f}\right)^2, \label{4c4}
\end{equation}
where $f$ is expressed by the following function:
\begin{equation}
f= \frac{1}{2} \log \left(4 R \left(\sqrt{A^4 \left(R^2-r^2\right)+4 r^4 R^2}+2 r^2 R\right)-A^4\right),
\end{equation}
 and $H$ and $K$ are integration constants. The density and pressure in Rastall theory are found to be
 \begin{widetext}
\begin{eqnarray}
\rho&=&r^{-2}\Bigg(f_1 \left(\frac{12 \alpha  r^2 R \left(\cos f - K \sin
f \right)}{f_2\left(K \cos f
+\sin f\right)}\right)+(4 \alpha -1) \left(f_1 -1\right)+ (\alpha
-1)
r \left(\frac{16 r^3}{A^4}-\frac{2 r}{R^2}\right)\Bigg),   \label{70a}\\
p&=&r^{-2}\Bigg(f_1 \Big(\frac{4 (3 \alpha -1) r^2 R \left(K \sin f -\cos
f \right)}{f_2 \left(K \cos f
+\sin f \right)}+\frac{2 \alpha  r^2 \left(A^4-8 r^2 R^2\right)}{A^4
\left(R^2-r^2\right)+4 r^4 R^2}\Big)+(4 \alpha -1)
\left(1-f_1\right)\Bigg),\label{70b}
\end{eqnarray}
\end{widetext}
where we have further set $f_1(r) =  \left(\frac{4
r^4}{A^4}-\frac{r^2}{R^2}+1\right)  $ and $f_2(r) = \sqrt{A^4 \left(R^2-r^2\right)+4 r^4 R^2} $. 
The  sound speed index, $dp/d\rho$, is given by 
\begin{widetext}
\begin{eqnarray}
\frac{dp}{d\rho} =&&\xi_{1}^{-1} \Big(-2 (2 \alpha +1) \left(K^2+1\right) R
f_2+\Big(4 R \left((1-8
\alpha ) K f_2 +2 (3 \alpha -1)
\left(K^2-1\right) r^2 R\right) \nonumber \\ &&-(3 \alpha -1) A^4
\left(K^2-1\right)\Big) \sin 2f +2
\Big(R \Big((1-8 \alpha ) K^2 f_2   \nonumber \\ &&+(8 \alpha -1) f_2 +8 (1-3
\alpha ) K r^2 R\Big) +(3 \alpha -1) A^4 K\Big) \cos 2f \Big), \label{71}\\
\xi_{1} =&& \Big(2 \Big((2 \alpha -5) \left(K^2+1\right)
R f_2 +\frac{1}{2} \Big(3
\alpha  K^2 \left(A^4-8 r^2 R^2\right)+4 (8 \alpha -5) K R f_2 \nonumber \\ &&-3 \alpha  \left(A^4-8 r^2
R^2\right)\Big) \sin 2f +\Big((8 \alpha -5) \left(K^2-1\right) R
f_2 -3 \alpha  A^4 K+24 \alpha K r^2 R^2\Big) \cos 2f \Big)\Big) .
\end{eqnarray}
\end{widetext}
The energy conditions are governed by the relations
\begin{widetext}
\begin{eqnarray}
\rho - p &=& \eta_{1}^{-1} \Big(4 f_1  \Big(-\Big(A^4 \left((3 \alpha -1) f_2 +(6 \alpha -1) K R \left(R^2-r^2\right)\right)+2 r^2 R^2 \left((3-8 \alpha ) f_2  +2 (6 \alpha -1) K r^2 R\right)\Big) \sin f \nonumber \\ && -K f_2  \left((3 \alpha -1) A^4+2 (3-8 \alpha ) r^2 R^2\right) \cos f +(6 \alpha -1) R f_2^2 \cos f \Big)\Big),  \label{72a}\\
\rho + p &=&  \eta_{2}^{-1}\Big(2 \Big(\Big(A^4 \left(f_2 +2 K r^2 R-2 K R^3\right)-8 r^2 R^2 \left(f_2 +K r^2 R\right)\Big) \sin f+\Big(8 r^2 R^2 \left(r^2 R-K f_2\right)\nonumber \\ &&+A^4 \left(K f_2 -2 r^2 R+2 R^3\right)\Big) \cos f\Big)\Big) , \label{72b}\\
\rho + 3p &=&  -\eta_{3}^{-1}\Big( \Big(\Big(3 A^4 \left((2 \alpha -1) K R \left(r^2-R^2\right)-\alpha  f_2 \right)+2 r^2 R^2 \left((8 \alpha +1) f_2 + 6 (1-2 \alpha ) K r^2 R\right)\Big) \sin f \nonumber \\ &&+\Big(2 r^2 R^2 \left((8 \alpha +1) K f_2 +6 (2 \alpha -1) r^2 R\right)-3 A^4 \left(\alpha  K f_2 +(2 \alpha -1) r^2 R-2 \alpha  R^3+R^3\right)\Big) \cos f \Big)\Big), \label{72c}
\end{eqnarray}
\end{widetext}
where we have defined new variables:
\begin{eqnarray}
&& \eta_{1}=f_2^{3} \left(K \cos f +\sin f \right),\nonumber \\ && \eta_{2}= A^4 R^2 f_2  \left(K \cos f +\sin f \right) ,\nonumber \\ && \eta_{3}= A^4 R^2 f_2  \left(K \cos f +\sin f\right).
\end{eqnarray}
For the purpose of the plots we have employed the parameter values $K = H = R = A = 1$. Observe that $K = 0$ corresponds to the partial solution obtained  by Tolman. We however consider the most general solution in the form of graphical plots. 

\begin{figure}
    \centering
 \includegraphics[width=0.3\textwidth, height=0.23\textheight]{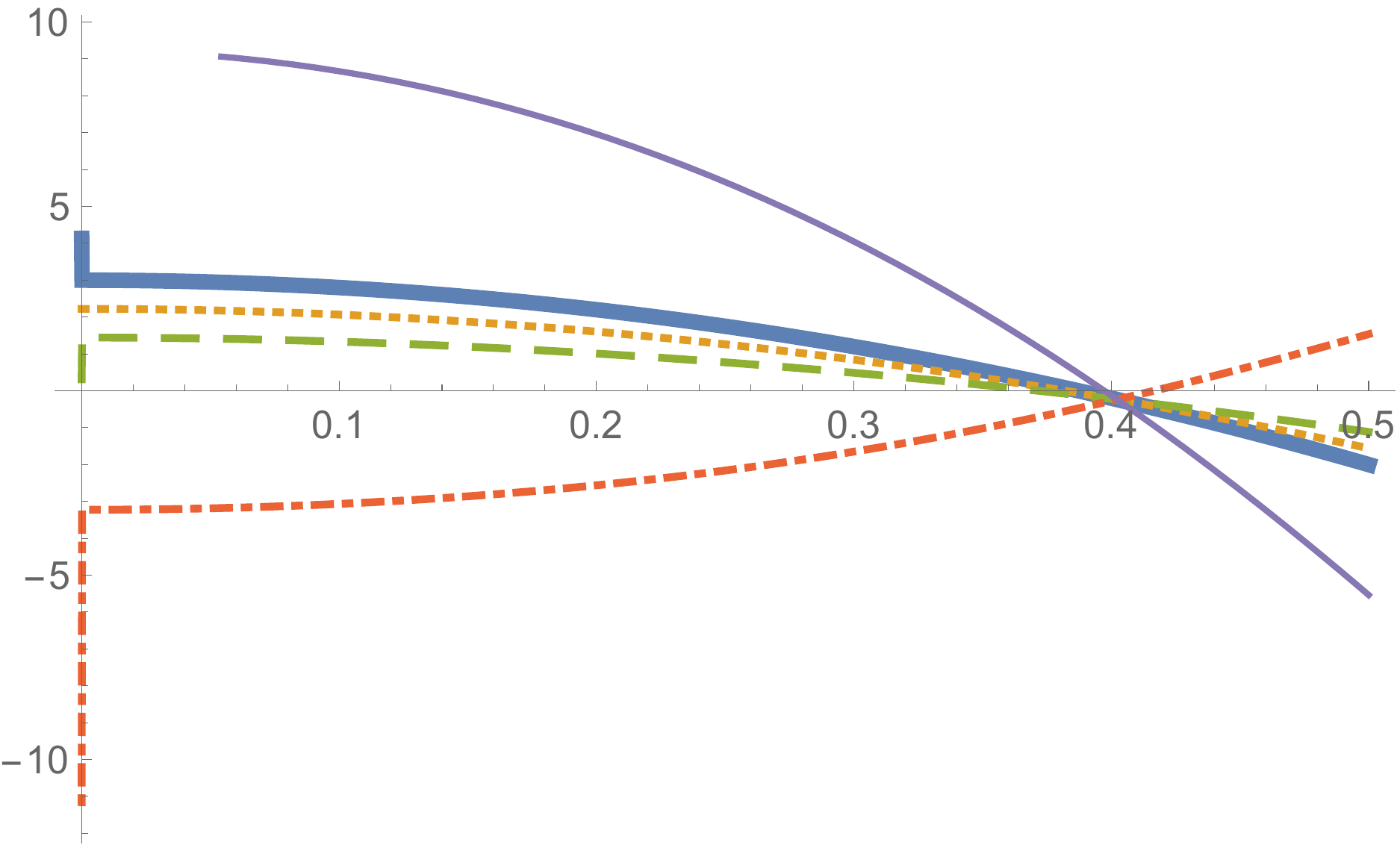}
    \caption{Energy density ($\rho$) versus radius ($r$): Tolman VII}
\end{figure}

\begin{figure}
    \centering
 \includegraphics[width=0.3\textwidth, height=0.23\textheight]{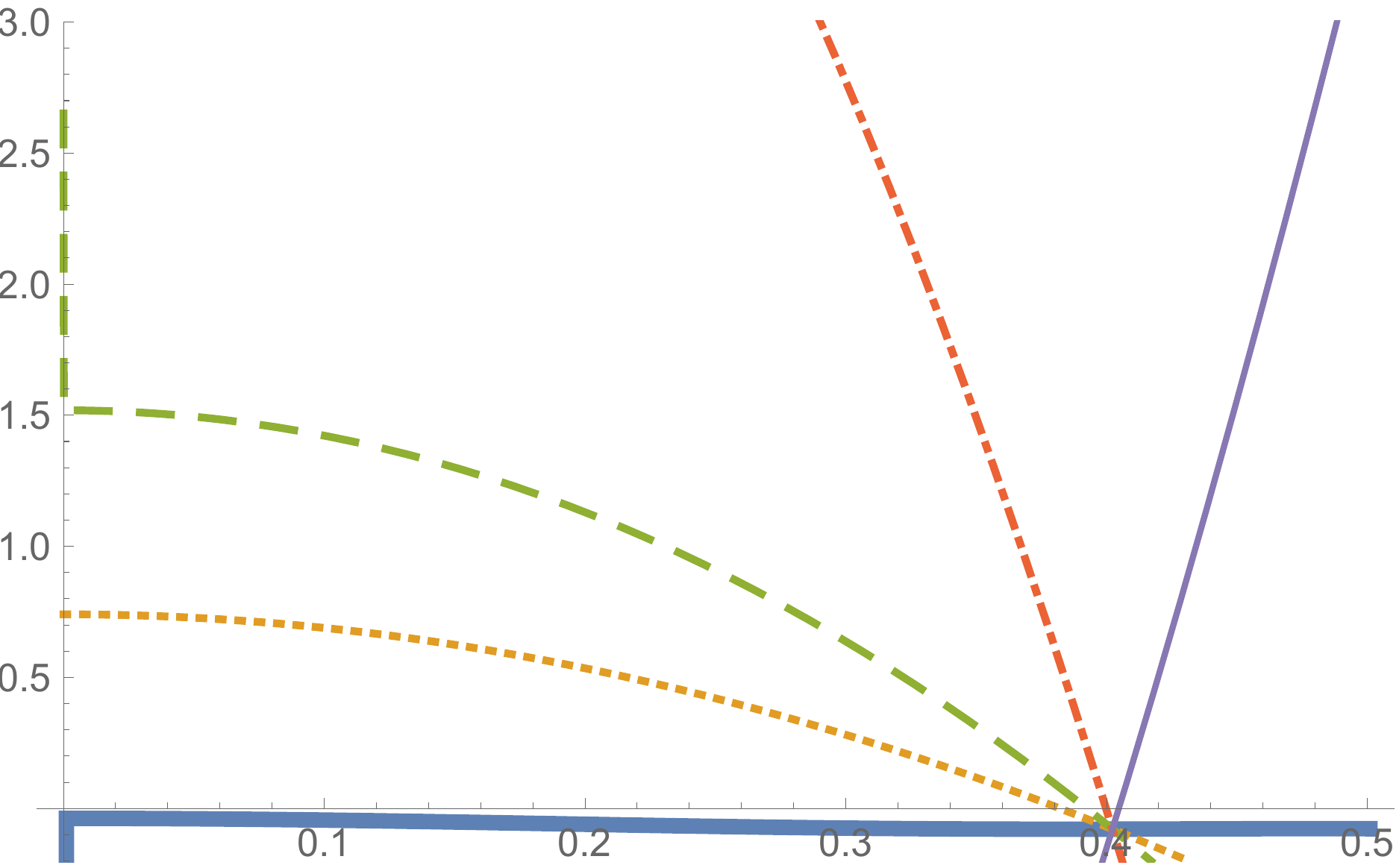}
    \caption{Pressure ($p$) versus radius ($r$): Tolman VII}
\end{figure}

\begin{figure}
    \centering
 \includegraphics[width=0.3\textwidth, height=0.23\textheight]{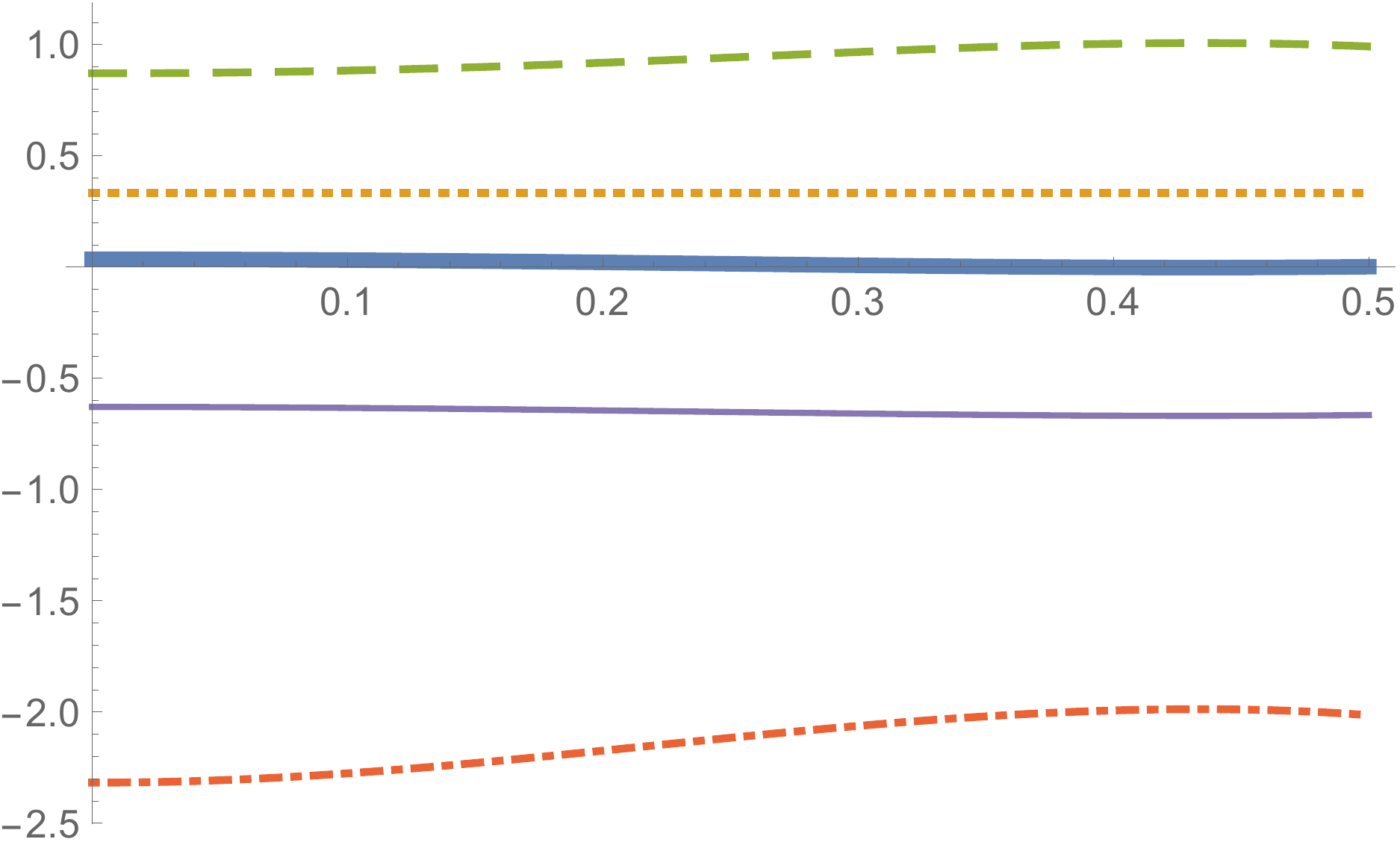}
    \caption{Sound speed versus radius ($r$): Tolman VII}
\end{figure}

\begin{figure}
    \centering
  \includegraphics[width=0.3\textwidth, height=0.23\textheight]{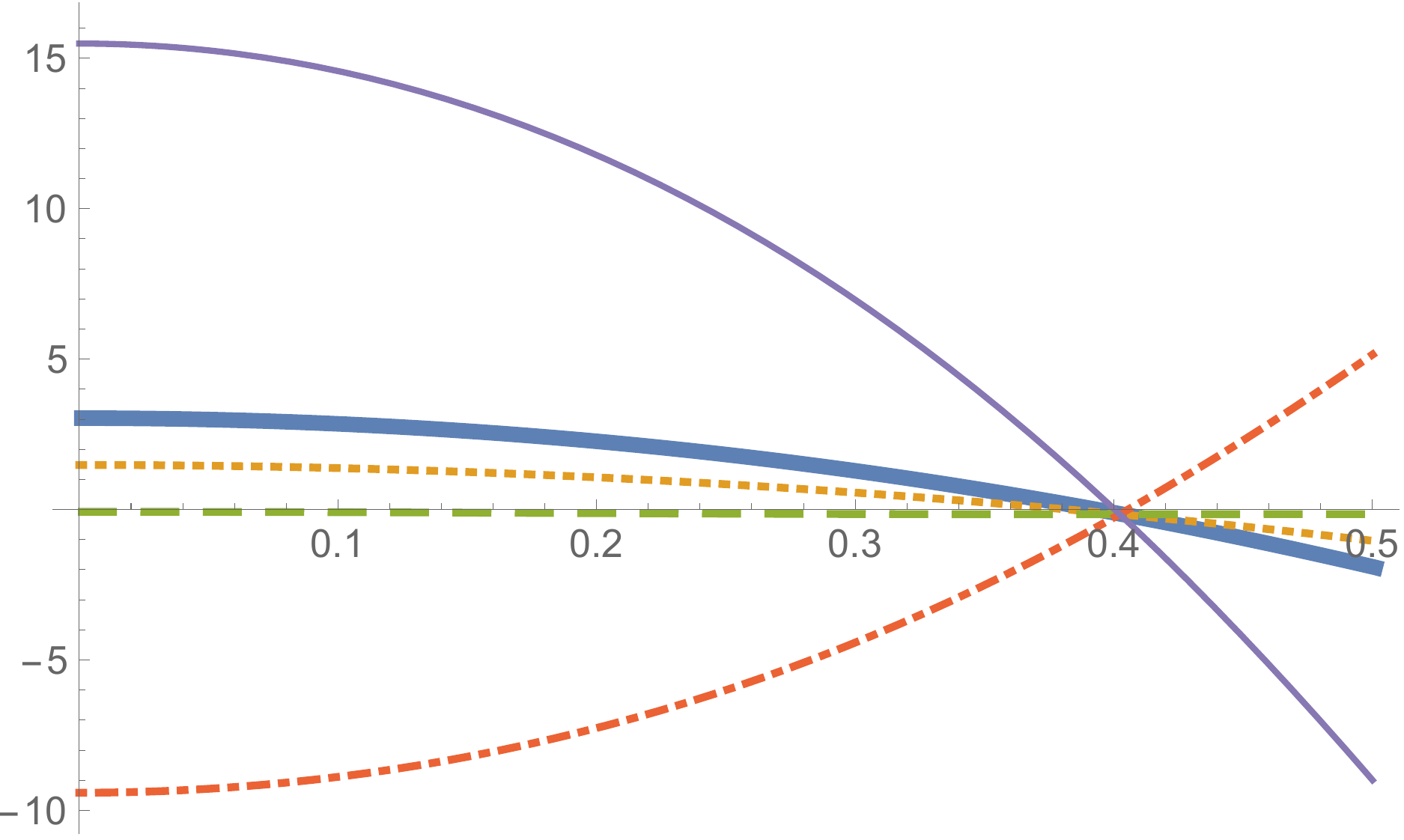}
    \caption{Weak energy condition versus radius ($r$): Tolman VII}
\end{figure}

\begin{figure}
    \centering
 \includegraphics[width=0.3\textwidth, height=0.23\textheight]{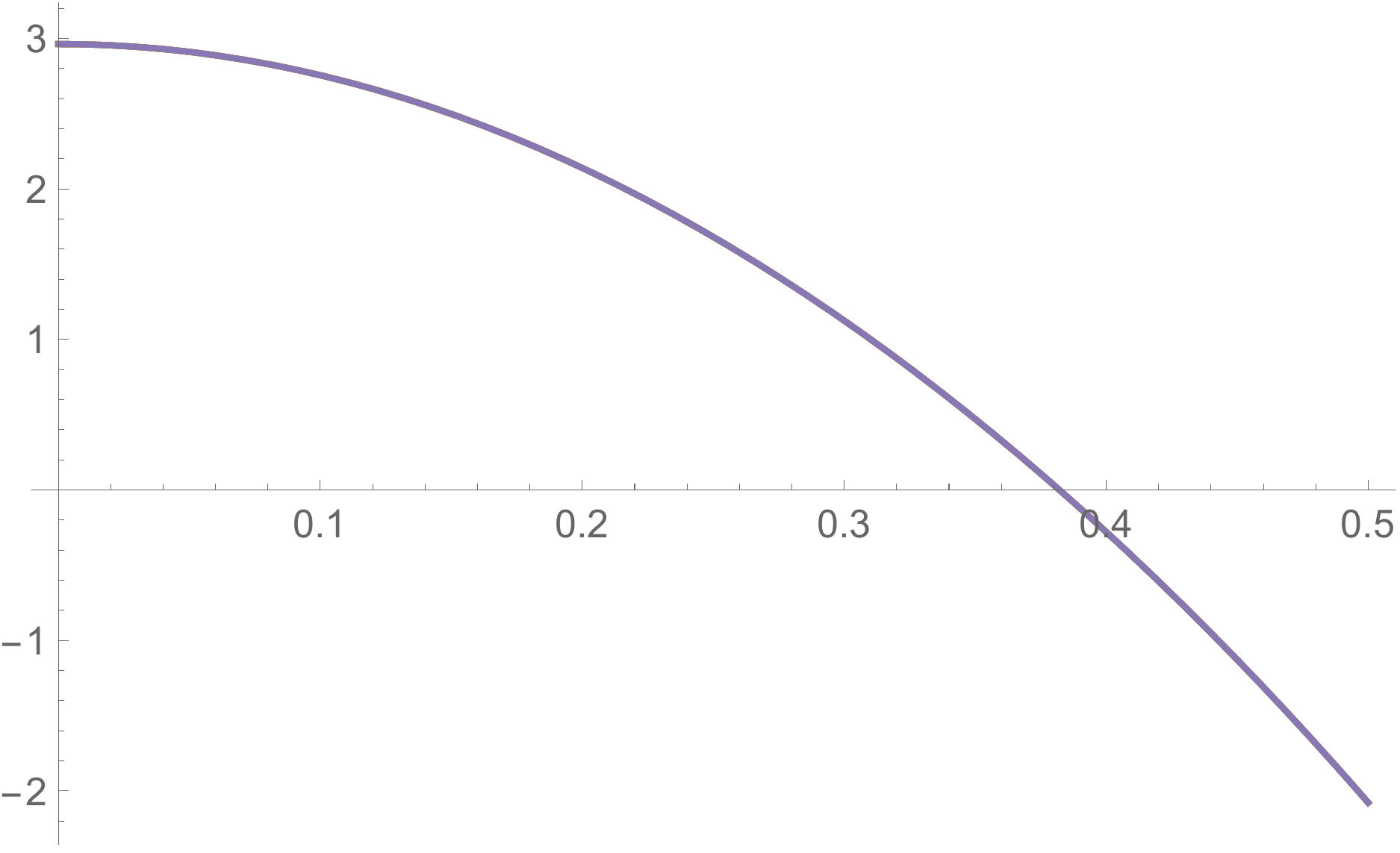}
    \caption{Strong energy condition versus radius ($r$): Tolman VII}
\end{figure}

\begin{figure}
    \centering
\includegraphics[width=0.3\textwidth, height=0.23\textheight]{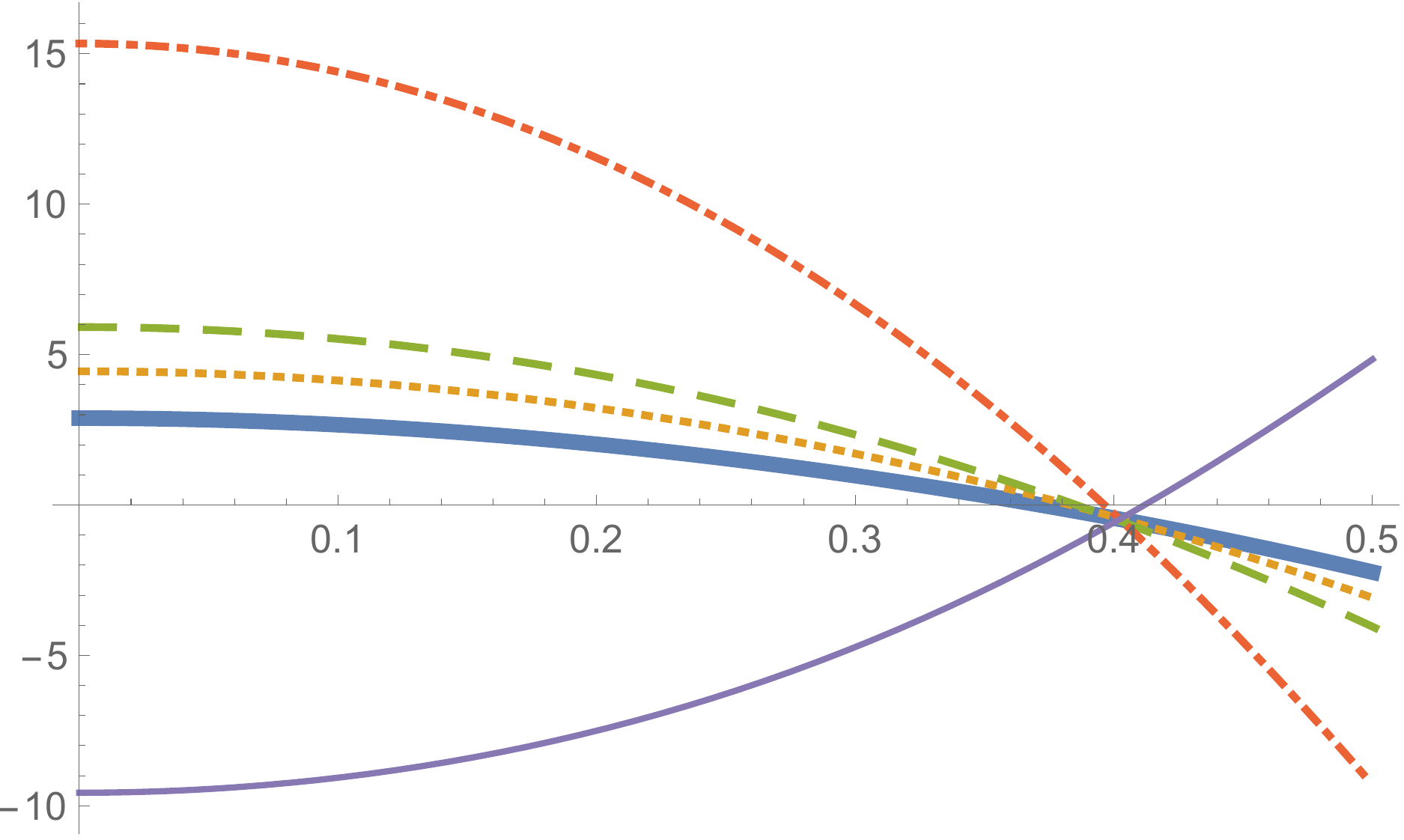}
    \caption{Dominant energy condition  versus radius ($r$): Tolman VII}
\end{figure}

This is amongst the least studied of the Tolman metrics given the complexity of the expressions. Moreover, Tolman presented only a special case of the general solution of the pressure isotropy equation. In the Einstein model ($\alpha = 0$), there is considerable deviation from a realistic fluid sphere. For example, the energy density (Fig 27) while being positive has a singularity at the centre $r=0$ and the pressure (Fig 28) is everywhere negative with no contact with the radial axis. This is physically unacceptable. On a positive note the energy conditions (Fig 30, 31, 32) are indeed satisfied. Once again it is a Rastall model ($\alpha = 0.25$) that  satisfies all the elementary physical requirements including a subluminal sound speed (Fig 29). The other Rastall curves display some pleasing characteristics but fail in some respects. It does not seem possible to display the mass profiles explicitly. Yet again the Rastall curve $\alpha = 0.25$ outshines the Einstein case. This further illustrates a distinction between the physics of models in general relativity and their counterpart Rastall theory.


\subsection{Tolman VIII metric}

With the postulated behaviour $ 
e^{-\lambda} =\text{const.}r^{-2b}e^{\nu}
$, the metric potentials are obtained in this case as  
\begin{eqnarray}
    e^{-\lambda}&=&\left(H-\frac{A}{r^n}- \frac{r^q}{F}\right)\,,\\ \nonumber
    e^{\nu}&=&B^2r^{2b} \left(H-\frac{A}{r^n}- \frac{r^q}{F}\right)
\end{eqnarray}
where, $H=\frac{2}{qn}$ and $A= (2m)^{n}$; with $q=a-b$, $n=a+2b-1$ and $F$ is a constant.

In the Rastall framework, the dynamics are governed by 
\begin{widetext}
\begin{eqnarray}
\rho =&&\Sigma^{-1}\Big( v_1\Big(2^{\frac{a^2+a+8}{3-a}} r \Big(2^{\frac{a+9}{a-3}} \alpha  a^6+2^{\frac{12}{a-3}} a^6-9\ 2^{\frac{a+9}{a-3}} \alpha  a^5+3\ 2^{\frac{12}{a-3}} a^5+2^{\frac{2 (a+3)}{a-3}} \alpha  a^4-11
\ 2^{\frac{a+9}{a-3}} a^4\nonumber\\&& +15\ 2^{\frac{4 a}{a-3}} \alpha  a^3- 3\, 2^{\frac{3 (a+1)}{a-3}} a^3-269\ 2^{\frac{a+9}{a-3}} \alpha  a^2+163\ 2^{\frac{12}{a-3}} a^2+105\ 2^{\frac{a+9}{a-3}} \alpha  a +49\ 2^{\frac{2 (a+3)}{a-3}} \alpha -99\ 2^{\frac{12}{a-3}} a\nonumber\\&& -35\ 2^{\frac{a+9}{a-3}}\Big) \left(\frac{m}{r}\right)^{-\frac{a^2+a-4}{a-3}} v_1^{-1} -2 (a+1) m v_1^{-1} \Big(2 a^5 (\alpha -1) \bar{v}_1 +a^4 \left(\alpha  \left(4-8 \bar{v}_1\right)+5 \bar{v}_1 -1\right)\nonumber\\&&+a^3 \left(-2 \alpha  \left(12\bar{v}_1 +11\right) +24 \bar{v}_1 +4\right)+a^2 \Big(20 \alpha  \left(6 \bar{v}_1 +1\right)-84 \bar{v}_1 +7\Big) -28 \alpha  \left(2 \bar{v}_1 +3\right)+a \Big(\alpha  \left(58-82 \bar{v}_1 \right)\nonumber\\&&+46 \left(\bar{v}_1-1\right)\Big)+35 \bar{v}_1 +48\Big)\Big)\Big), \label{80a}\\
p=&&\Sigma^{-1}\Big(v_1 \Big(2 (a+1) m v_1^{-1} \Big(2 a^5 \alpha  \bar{v}_1 +a^4 \left(\alpha  \left(4-8 \bar{v}_1\right)+ \bar{v}_1-1\right)+a^3 \left(6-2 \alpha  \left(12 \bar{v}_1   
+11\right)\right) \nonumber\\&& +a^2 \left(20 \alpha  \left(6 \bar{v}_1 +1\right)-3 \left(4 \bar{v}_1 + 3\right)\right) -28 \alpha  \left(2 \bar{v}_1 +3\right)+a \left(\alpha  \left(58-82 \bar{v}_1\right)+12 \bar{v}_1 -4\Big) +7 \bar{v}_1 +12\right)\nonumber\\&& -2^{\frac{a^2+a+8}{3-a}} r \Big(2^{\frac{a+9}{a-3}} \alpha  a^6-2^{\frac{12}{a-3}} a^6-9\ 2^{\frac{a+9}{a-3}} \alpha  a^5+5\ 2^{\frac{12}{a-3}} a^5+2^{\frac{2 (a+3)}{a-3}} \alpha  a^4 +3\ 2^{\frac{a+9}{a-3}} a^4+15\ 2^{\frac{4 a}{a-3}} \alpha  a^3-9\ 2^{\frac{3 (a+1)}{a-3}} a^3\nonumber\\&& -269\ 2^{\frac{a+9}{a-3}} \alpha  a^2+125\ 2^{\frac{12}{a-3}} a^2+105\ 2^{\frac{a+9}{a-3}} \alpha  a  +49\ 2^{\frac{2 (a+3)}{a-3}} \alpha -37\ 2^{\frac{12}{a-3}} a-21\ 2^{\frac{a+9}{a-3}}\Big) \left(\frac{m}{r}\right)^{-\frac{a^2+a-4}{a-3}} v_1^{-1}\Big)\Big), \label{80b}
\end{eqnarray}
\end{widetext}
where $\Sigma=\left(2 (a-3) \left(a^2-2 a-1\right) \left(a^2+2 a-7\right) m r^2 \right)$, and  we have redefined  $v_1 = \left(\frac{r}{R}\right)^{\frac{(a-2) (a+1)}{a-3}}$ and $\bar{v}_1 = \left(\frac{r}{R}\right)^{a+\frac{(a-2) (a+1)}{a-3}}$. 
The sound speed index, energy conditions and mass are all obtainable but omitted as they are lengthy. These have been plotted and the parameter values $a =1.5$, $m=1$, $R=2$ and $B=1$ have been used.

\begin{figure}
    \centering
 \includegraphics[width=0.3\textwidth, height=0.23\textheight]{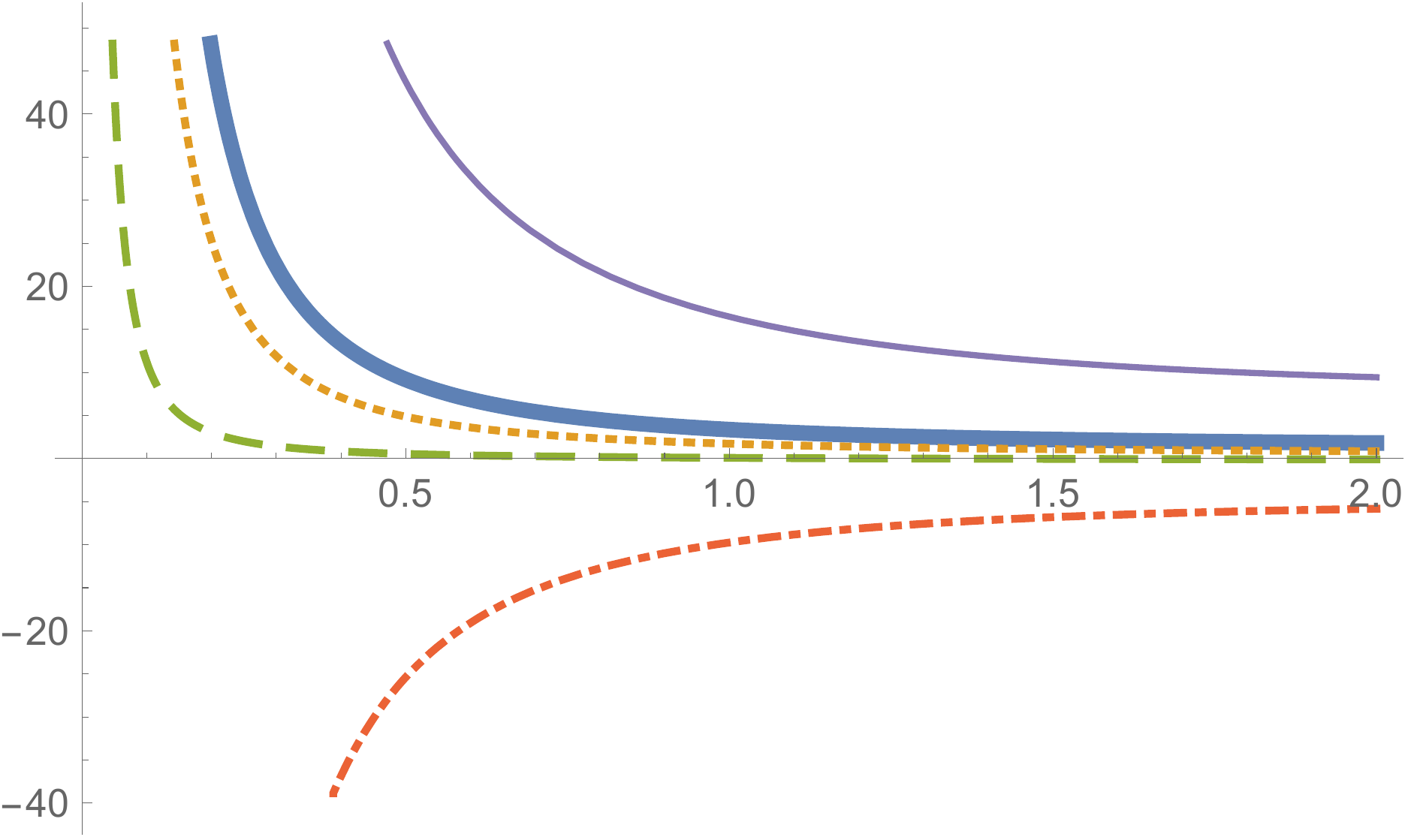}
    \caption{Energy density ($\rho$) versus radius ($r$): Tolman VIII}
\end{figure}

\begin{figure}
    \centering
 \includegraphics[width=0.3\textwidth, height=0.23\textheight]{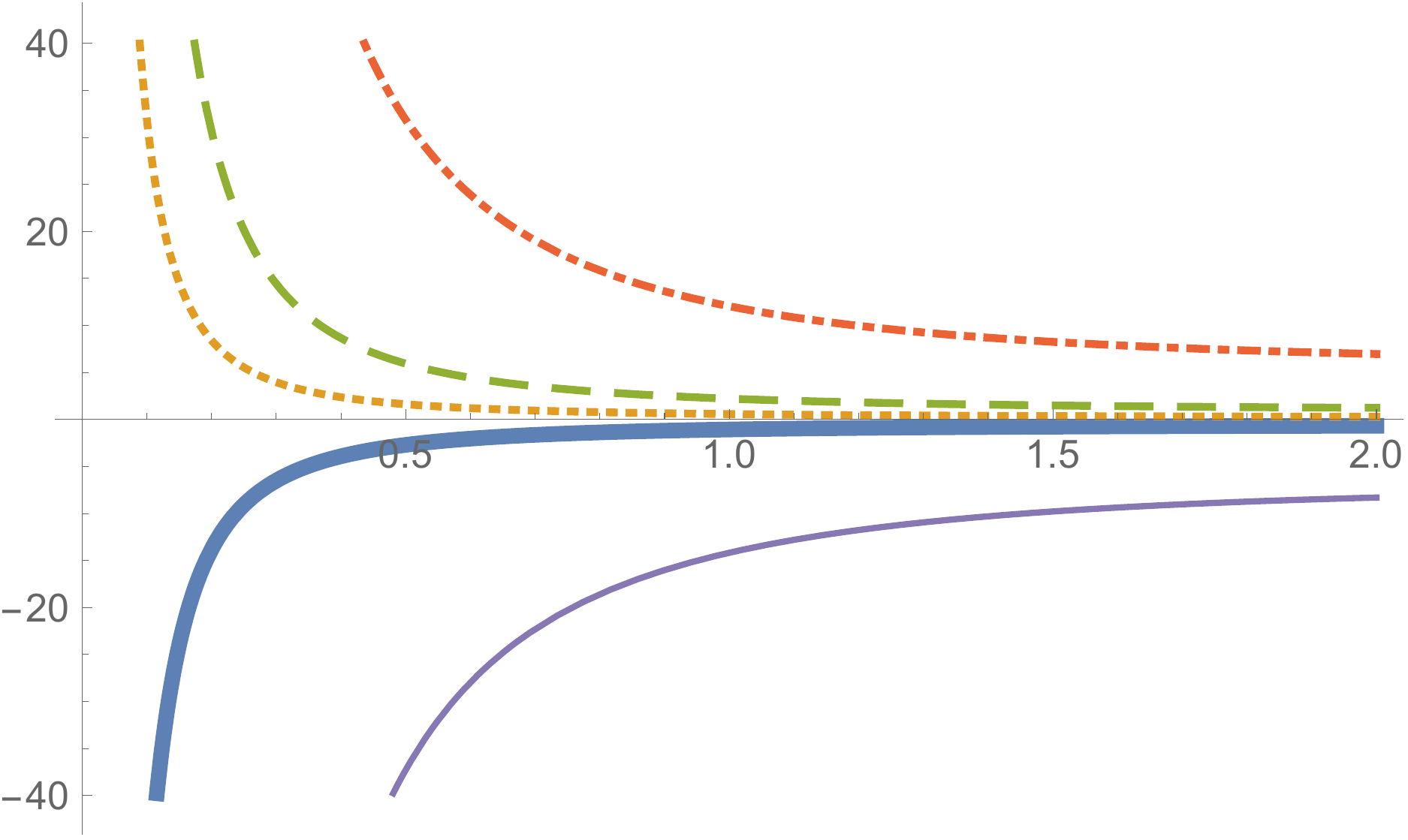}
    \caption{Pressure ($p$) versus radius ($r$): Tolman VIII}
\end{figure}

\begin{figure}
    \centering
 \includegraphics[width=0.3\textwidth, height=0.23\textheight]{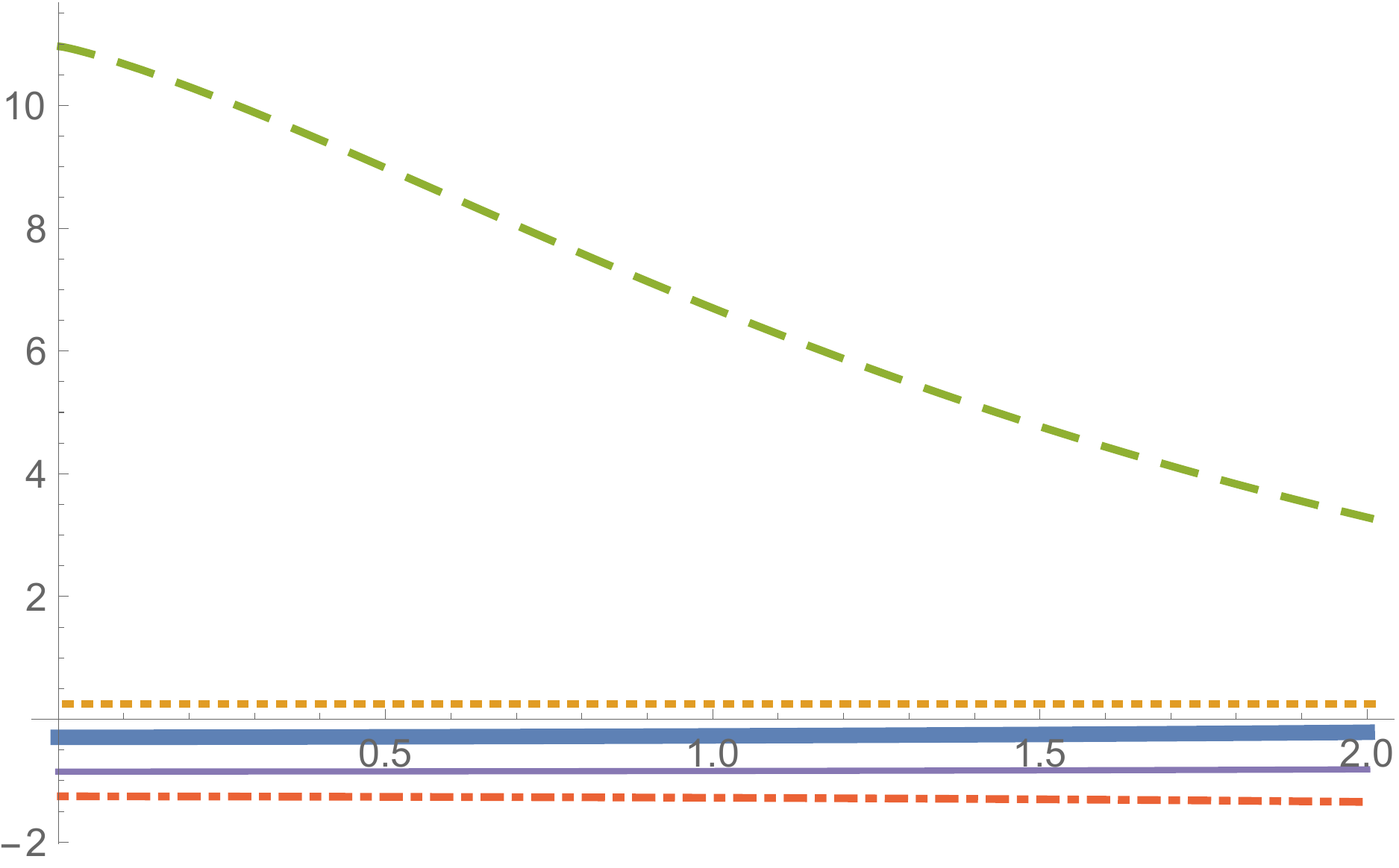}
    \caption{Sound speed versus radius ($r$): Tolman VIII}
\end{figure}

\begin{figure}
    \centering
  \includegraphics[width=0.3\textwidth, height=0.23\textheight]{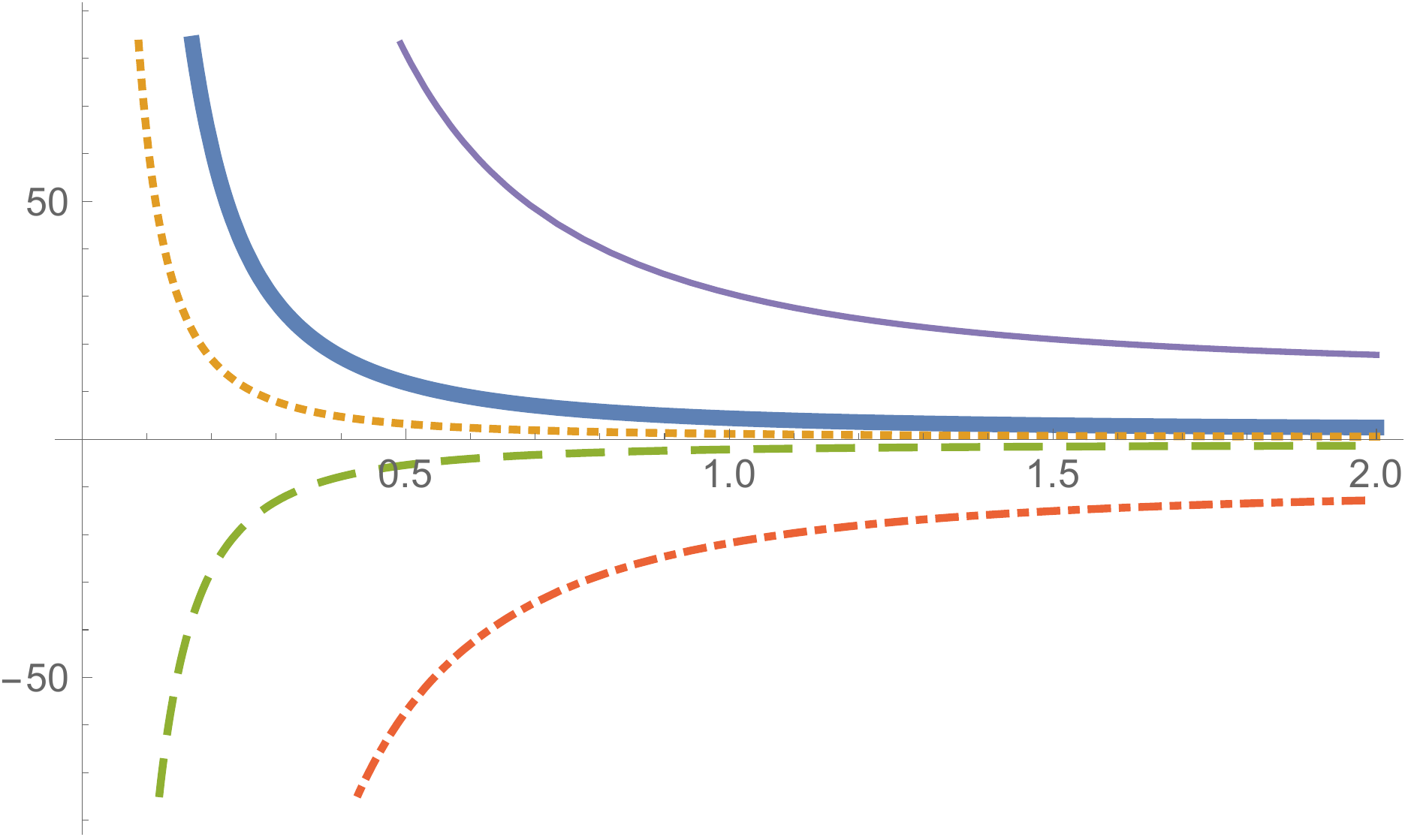}
    \caption{Weak energy condition versus radius ($r$): Tolman VIII}
\end{figure}

\begin{figure}
    \centering
 \includegraphics[width=0.3\textwidth, height=0.23\textheight]{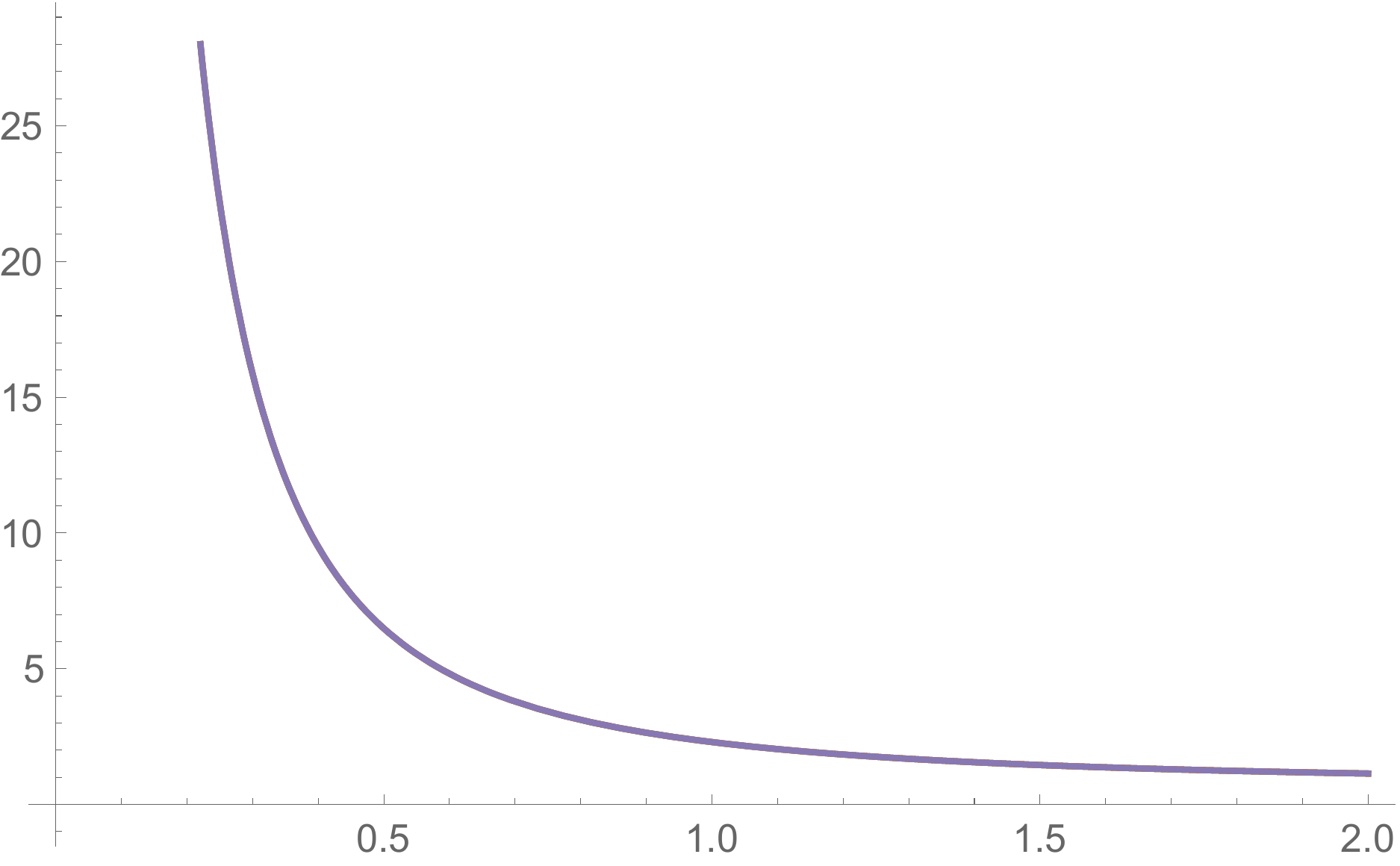}
    \caption{Strong energy condition versus radius ($r$): Tolman VIII}
\end{figure}

\begin{figure}
    \centering
 \includegraphics[width=0.3\textwidth, height=0.23\textheight]{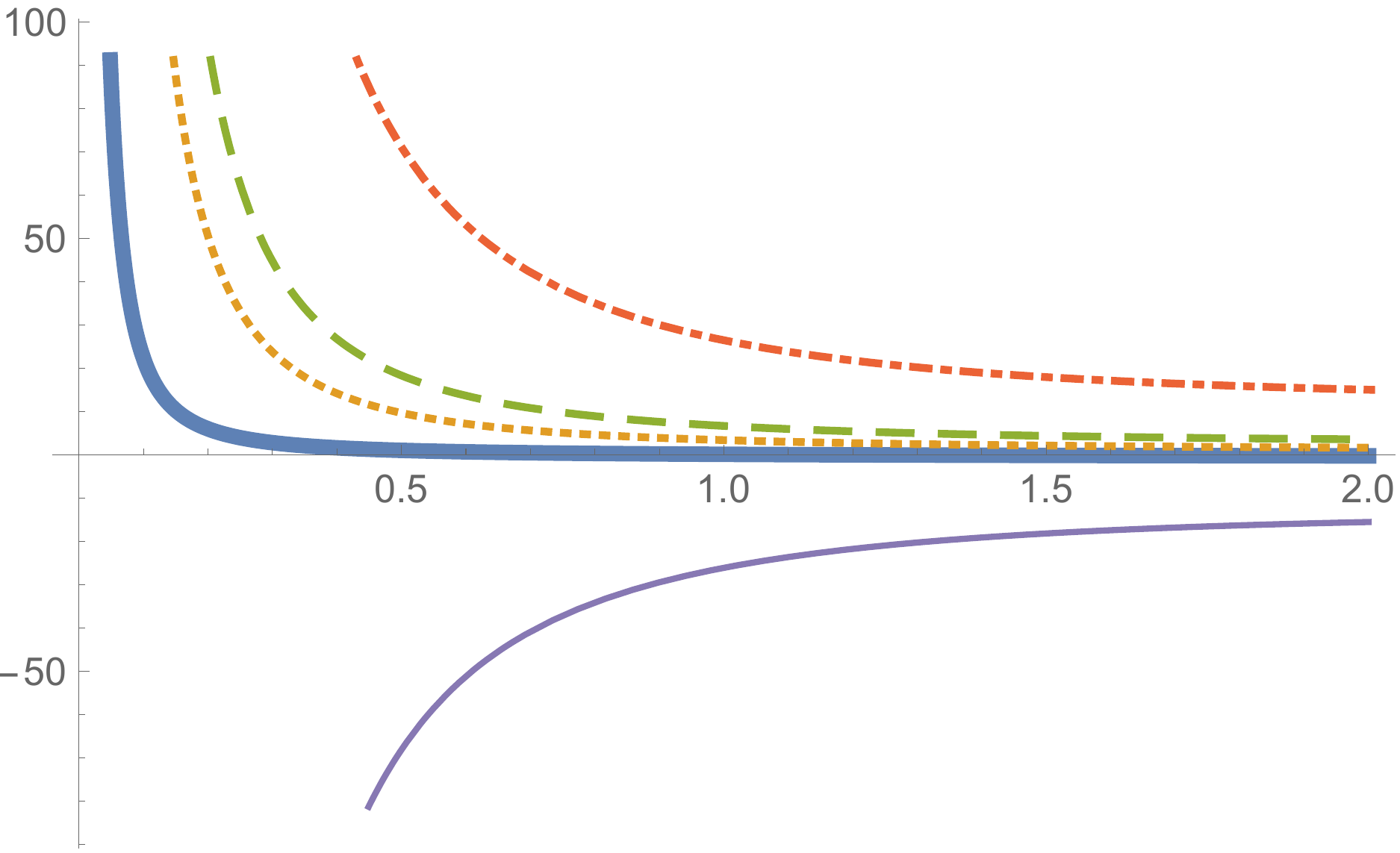}
    \caption{Dominant energy condition  versus radius ($r$): Tolman VIII}
\end{figure}

\begin{figure}
    \centering
 \includegraphics[width=0.3\textwidth, height=0.23\textheight]{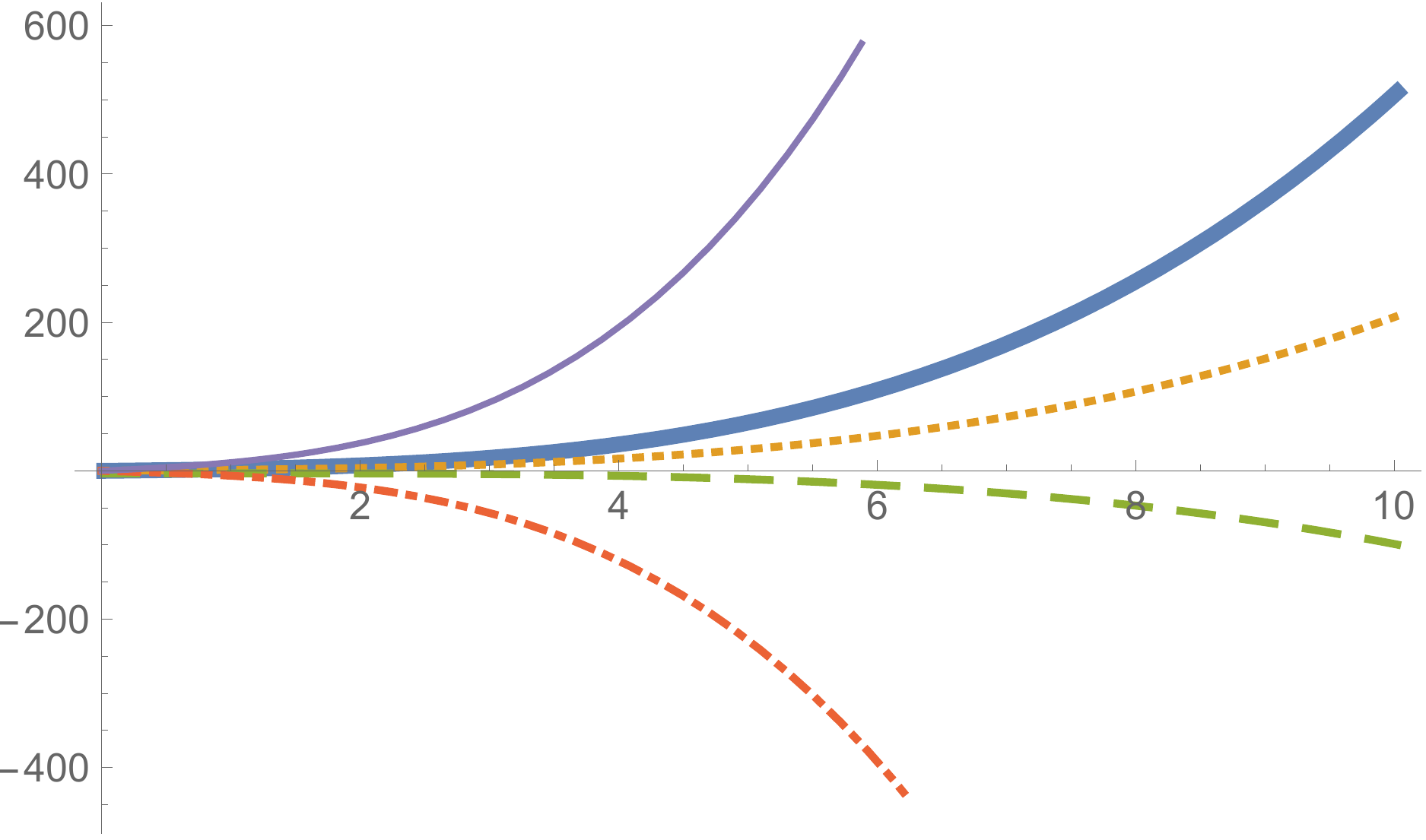}
    \caption{Mass   versus radius ($r$): Tolman VIII}
\end{figure}

This is another case with unwieldy expressions. However with the help of the mathematical tools, a graphical study is possible. In this case, the Einstein version shows a positive density profile (Fig 33), a negative pressure (Fig 34) and a violation of causality (Fig 35). All the energy conditions (Fig 36, 37, 38) appear to be satisfied. The mass profiles (Fig 39) are well behaved in all but one Rastall case.  The Rastall case $\alpha = 0.25$ again satisfies all the physical requirements. What is also interesting to note is the changes in the profiles of the graphs for different choices of the Rastall parameter. This should provide conclusive evidence that the Rastall gravity theory is not equivalent to the Einstein theory.  

\section{Conclusion}

In this work, we analyze the behaviour of the Tolman metrics in the Rastall theory. The study shows incontrovertibly that the equivalence between the Rastall framework and the  standard theory of Einstein as purported recently in Ref.\cite{Visser} is confined to the identical geometry but not the same physics. What is conceded is that the equation of pressure isotropy remains the same as in Einstein theory. Accordingly any of the metrics satisfying Einstein's field equations may be used to generate density and pressure profiles in the new Rastall paradigm.  Remarkably where the Einstein theory fails to satisfy the elementary requirements for physical plausibility, the Rastall version succeeds. For example, the Tolman VII model depicts a negative pressure profile with a central singularity while  the Rastall version is finite at the centre, decreases monotonically outwards and vanishes for a certain radial value. This is expected for realistic models.  Additionally it has been demonstrated that the Rastall case $\alpha = 0.25$ satisfies all the reasonably physical requirement in all Tolman models and succeeds even in Tolman models that are considered unphysical in the Einstein theory. There are thus indications that the Rastall theory may be promising as a theory of gravitation as it supports astrophysical objects with pleasing behaviour. The non-conservation of energy-momentum has been explained by Rastall as being an artifact of spacetime curvature. Nevertheless, we observe that the $\alpha = 0.25$ case indeed satisfies the weak, strong and dominant energy conditions although this does not mean that energy is conserved. Moreover, a surface of vanishing pressure exists in most models and the causality principle is respected in all cases. The mass profiles conform to expectations. In these respects we find that the Rastall models display more pleasing physical contributions than their Einstein counterparts. It is clear from this empirical analysis that the evidence for the physical equivalence of Rastall theory and general relativity is not present. Instead the opposite appears to be true. The Rastall parameter offers a mathematical handle to correct the deficiencies in the standard theory.  For these reasons it would not be reasonable to discount such phenomenological ideas. The jury is still out on the true theory of gravity. This study shows that astrophysical models that are physically viable may be constructed with ease in this framework. The import of this is that the reliability of this framework to provide answers to deeper questions such as explaining the late-time accelerated expansion of the universe without resorting to invocations of exotic matter or dark matter fields, is strengthened.

\textbf{Acknowledgments}:  
AB thanks the organizers of University of Kwazulu-Natal for financial support. SH thanks the National Research Foundation of South Africa for support.

\end{document}